\documentclass[11pt,a4wide]{article}
\pdfoutput=1

\usepackage[shortlabels]{enumitem}
\usepackage{jheppub}
\usepackage{hyperref}

\usepackage{makecell} 
\usepackage[english]{babel}
\usepackage[babel]{csquotes}
\usepackage{amsfonts}
\usepackage{array, tabularx}
\usepackage{mathtools}
\usepackage{amsthm}
\usepackage{url}
\usepackage{multicol, multirow}
\usepackage{emptypage}
\usepackage{verbatim}
\usepackage{dsfont}
\usepackage{hyperref}
\usepackage{float}
\usepackage{amsmath}
\usepackage{amssymb,tikz}
\usepackage{verbatim}
\usepackage{xspace}

\newcommand{\1}{{\mathds{1}}}
\renewcommand{\O}{{\mathcal O}}

\newcommand{\de}{{\partial}}

\newcommand{\eps}{{\varepsilon}}

\usepackage{tensor}

\newcommand{\Z}{\mathbb Z}

\allowdisplaybreaks[2]

\usepackage[font=small,format=hang,labelfont={sf,bf}]{caption}

\makeatother

\DeclareMathOperator{\Tr}{Tr}

\title{Spectrum continuity and level repulsion: the Ising CFT from infinitesimal to finite $\boldsymbol\varepsilon$}

\author[1]{Johan Henriksson,}
\author[1,2,3]{Stefanos R. Kousvos}
\author[4]{\& Marten Reehorst}
\affiliation[1]{Department of Physics, University of Pisa and INFN, \\Largo Pontecorvo 3, I-56127 Pisa, Italy}
\affiliation[2]{Department of Physics, University of Crete, Heraklion GR-70013, Greece}
\affiliation[3]{Institute of Theoretical and Computational Physics (ITCP), Department of Physics,
University of Crete, 70013 Heraklion, Greece}
\affiliation[4]{CPHT, CNRS, \'Ecole Polytechnique, Institut Polytechnique de Paris,\\Route de Saclay, 91128 Palaiseau, France}

\emailAdd{johan.henriksson@df.unipi.it}
\emailAdd{stefanos.kousvos@df.unipi.it}
\emailAdd{marten.reehorst@polytechnique.edu}

\newcommand{\N}{\mathcal{N}}

\abstract{
Using numerical conformal bootstrap technology we perform a non-perturbative study of the Ising CFT and its spectrum from infinitesimal to finite values of $\varepsilon=4-d$. Exploiting the recent navigator bootstrap method in conjunction with the extremal functional method, we test various qualitative and quantitative features of the $\varepsilon$-expansion. 
We follow the scaling dimensions of numerous operators from the perturbatively controlled regime to finite coupling.
We do this for $\mathbb Z_2$-even operators up to spin 12 and for $\mathbb Z_2$-odd operators up to spin 6 and find a good matching with perturbation theory.
In the finite coupling regime we observe two operators whose dimensions approach each other and then repel, a phenomenon known as level repulsion and which can be analyzed via operator mixing. 
Our work improves on previous studies in both increased precision and the number of operators studied, and is the first to observe level repulsion in the conformal bootstrap. 

}

\begin{document}
\maketitle


\section{Introduction}

The past decades have brought various advances in both perturbative and non-pertubative methods for studying conformal field theories. On the perturbative side, there is the $\varepsilon$-expansion which started with Wilson and Fisher \cite{Wilson:1971dc}\footnote{See also  \cite{Wilson:1973jj,Kleinert2001,Pelissetto:2000ek}
for an incomplete list of useful references.} and recently led to the computation of anomalous dimensions up to order $\eps^7$ and $\eps^8$ \cite{Schnetz:2016fhy,SchnetzUnp}. On the other hand there is the non-perturbative numerical conformal bootstrap, whose flagship application has been the most precise determination of the critical exponents in the 3d Ising model \cite{Kos:2016ysd}. With these tools in hand, the time is ripe to perform a precision study on the relation between the perturbative and non-perturbative descriptions. 
We also probe the conjecture that a continuous family of conformal field theories (CFTs) exists for any dimension between $3 \leqslant d \leqslant 4$ connecting the 3d Ising CFT to the Gaussian theory in 4d.\footnote{This assumption has been implicit in a lot of work, see e.g. \cite{LeGuillou:1987ph}, and was formulated as ``spectrum continuity'' in \cite{Hogervorst:2015akt}. The family is further conjectured to exist all the way down to $d=2$ where it is believed to merge with the 2d Ising model originally solved in \cite{Onsager:1943jn}.
}

In addition, we will test for the existence of inherently finite-coupling effects such as level repulsion, known from quantum mechanics as the von Neumann--Wigner non-crossing rule \cite{vonNeumann1929} or avoided crossing.
In the context of CFTs, level repulsion concerning eigenvalues of the dilatation operator was discussed in \cite{Korchemsky:2015cyx}, who noted that if two operators have dimensions that get close to each other, when varying a coupling, their dimensions can be approximated by the eigenvalues of a two-state ``Hamiltonian''
\begin{equation}
\label{eq:mixinggenIntro}
\mathbb D= \begin{pmatrix}
\Delta_1(g) & x(g) \\ x(g) & \Delta_2(g)
\end{pmatrix}.
\end{equation}
The approximation holds for small $x$, in the region where $|\Delta_1-\Delta_2|$ is comparable to $x$. An explicit example satisfying these assumptions was given in the case of the large $N$ expansion of $\mathcal N=4$ super-Yang--Mills (SYM) theory. One of the goals of our paper is to investigate if \eqref{eq:mixinggenIntro} is a good model also in the Ising CFT, where the mentioned assumptions are not guaranteed to hold. In our context, the role of $g$ is played by the spacetime dimension $d$, or equivalently $\eps=4-d$.

We investigate the above questions by studying the Ising CFT in fractional dimensions via the conformal bootstrap. Although the $\varepsilon$-expansion predicts that the conjectured family of CFTs is non-unitary for non-integer $d$ \cite{Hogervorst:2015akt}, it was found that the numerical conformal bootstrap is in practice insensitive to this non-unitarity \cite{El-Showk:2013nia,Cappelli:2018vir}. This is likely because the operators responsible for the non-unitarity occur at large scaling dimensions \cite{Hogervorst:2015akt}, to which the bootstrap is less sensitive.

We find a continuous family of allowed islands in parameter space for spacetime dimensions in the range $3 \leqslant d \leqslant 4$ (as well as below $d=3$). Inside these islands we extract the approximate solution to crossing via the extremal functional method (EFM) \cite{El-Showk:2012vjm,El-Showk:2014dwa,Komargodski:2016auf,Simmons-Duffin:2016wlq,El-Showk:2016mxr}. We then compare the resulting spectrum against known perturbative results obtained in the $\varepsilon$-expansion for $\phi^4$ theory.
In addition to the anomalous dimensions related to the usual critical exponents (such as $\beta = \frac{\Delta_{\phi}}{d-\Delta_{\phi^2}}$) we also compare against perturbative predictions for a much larger part of the spectrum. These predictions were computed in \cite{Kehrein:1992fn,Kehrein:1994ff,Hogervorst:2015akt,Henriksson:2022rnm}.

Let us briefly summarize the numerical conformal bootstrap method. For an overview of results see the review papers \cite{Poland:2018epd} and \cite{Poland:2022qrs}. The flagship application of the numerical conformal bootstrap is to find bounds with rigorous error bars on conformal data -- operator dimensions and operator product expansion (OPE) coefficients -- that enter in the four-point function of primary operators. These bounds are found by systematically probing parameter space for self consistency (defined as crossing symmetry plus unitarity). The demand of self consistency can be phrased in terms of the existence, or not, of a functional with specific properties (positive action on all operator contributions). If a functional exists for a specific choice of conformal data, one concludes that this set of data cannot correspond to a self consistent CFT, as defined above. In addition, one may find complementary non-rigorous approximations of the operator spectrum by looking for minima of the aforementioned functional.

The Ising CFT has been subject to several previous non-perturbative numerical bootstrap studies, with the goal of determining its operator spectrum and other CFT-data, or as a benchmark target for new techniques. Important for our work is \cite{Simmons-Duffin:2016wlq}, where a precision estimate was given for dimensions and OPE coefficients of numerous operators in the low dimensional operators in the 3d Ising model, obtained via the EFM. 
That paper also verified that the leading twist families behave as expected from analytic computations in the light-cone limit \cite{Fitzpatrick:2012yx,Komargodski:2012ek,Caron-Huot:2017vep}. Part of this spectrum was further studied in \cite{Reehorst:2021hmp} where rigorous error bars were provided for a larger set of CFT data involving 13 quantities. 
Among other bootstrap approaches is the truncation method of \cite{Gliozzi:2013ysa} which was applied to the 3d Ising model in \cite{Gliozzi:2014jsa}. In this approach instead of rigorously carving out parameter space,
a truncated set of crossing equations 
is solved by assuming only a small finite number of exchanged operators. 
Another similar approach is to minimize the error that arises from truncating the crossing equation.
This was recently implement in \cite{Kantor:2021jpz,Kantor:2021kbx} using machine learning, although it only explicitly investigated the case of $D=2$. See also \cite{Laio:2022ayq} for an alternative of this idea using Monte Carlo. Finally we should mention the recent hybrid numeric-analytic bootstrap proposed by \cite{Su:2022xnj}. Here the numerical bootstrap is supplemented by information on the large spin operators coming from the analytic light-cone bootstrap. While not strictly rigorous this was found to greatly improve the constraining power of the numerical bootstrap. 

Most pertinent to this study, \cite{El-Showk:2013nia,Cappelli:2018vir} previously used the numerical conformal bootstrap to study the Ising CFT at finite fractional $\eps$ by following the position of a kink in the $(\Delta_\sigma,\Delta_\epsilon)$ exclusion plot across various values of $d$ between 2 and 4. 
The approach we take is similar, we also work in fractional values of $d$ but instead of following the position of a kink in parameter space, we follow the position of an isolated island obtained using a mixed correlator bootstrap setup.\footnote{Such a mixed correlator bootstrap setup has previously been used to find islands at $3$, $3.25$, $3.5$ and $3.75$ space time dimensions \cite{Behan:2016dtz}. We will study more values of $d$ at higher derivative order and we will additionally be extracting the extremal spectrum (which cannot reliably be done at a lower derivative order).} We make use of the recently discovered navigator method \cite{Reehorst:2021ykw} to systematically locate small unitary islands in intermediate spacetime dimensions $2.6\leqslant d<4$,\footnote{The original range considered was $3\leqslant d<4$, however we extended the lower limit when we observed that the targeted level repulsion effect would happen below rather than above three dimensions.} in steps of $0.1$. Each island is isolated in terms of the variables ($\Delta_\sigma$, $\Delta_\epsilon$, $\theta$). Here $\Delta_\sigma$ and $\Delta_\epsilon$ are the scaling dimensions of the only relevant $\mathbb{Z}_2$-odd and $\mathbb{Z}_2$-even operators and $\theta=\arctan(\lambda_{\epsilon\epsilon\epsilon}/ \lambda_{\sigma \sigma \epsilon })$ is the OPE coefficient angle. Within these islands we obtain the spectrum of operators in the theory via the EFM. 

In order to obtain a CFT spectrum by the EFM, one has to fix specific values of for dimensions of the external operators and choose a parameter to extremize. For the latter there are in principle infinitely many choice, e.g. we can maximize any OPE coefficients or superposition of OPE coefficients. In \cite{Simmons-Duffin:2016wlq} a sampling over $20$ points was performed and minimization of the central charge was chosen as the direction of extremization. This gave an estimate of all low-lying CFT data. 
The standard deviation within this sample population was then interpreted as an estimate of the error on these values. The rigorous bounds of \cite{Reehorst:2021hmp} were found to closely match the intervals given by the error bars of the non-rigorous estimates.
Since the spread in the CFT data under the  mentioned sampling was found to be small, we avoid the computationally costly process of sampling and study only a single spectrum spectrum for each dimension. 
In order to have a unique well-defined selection criterion we examine only the spectrum at the minimum of the navigator function.\footnote{To be precise, we minimize $C_T$ at the external operator values corresponding to the navigator minimum.} 
This point can be seen as the ``most allowed point'' in the sense that it is the most probable to remain un-excluded when increasing the number of constraints. 
We hypothesize that the spectrum at this point gives a good estimate of the true spectrum of the underlying CFT, see some comparisons in section~\ref{sec:systematicErr}.  

Indeed, we find that spectra close to the navigator minimum match very well with the estimates from the $\varepsilon$-expansion for small $\varepsilon$. Surprisingly, in some cases even order $\varepsilon^1$ un-resummed predictions for scaling dimensions compare favourably with the non-perturbative $d=3$ bootstrap results. For the case of $\mathbb{Z}_2$-even scalars we find a clear example of level repulsion, involving the operators with third and fourth smallest scaling dimensions. We also investigate which operators present in the $\eps$-expansion are first to go undetected in the EFM. We find that whether an operator is detected or not is not just determined by the dimension of the operator. Looking at e.g. the $\mathbb{Z}_2$-even spin-$0$ sector we see that higher dimensional operators with a larger OPE coefficient are detected while lower dimensional operators with smaller OPE coefficients are overlooked. 

To denote operators we will mostly follow the numerical bootstrap conventions and denote $\mathbb{Z}_2$-even scalars by $\epsilon, \epsilon', \ldots$ and $\mathbb{Z}_2$-odd scalars by $\sigma, \sigma', \ldots$.
However, when comparing the numerical data with the $\varepsilon$-expansion we will also use the notation that is conventional in that context, $\sigma=\phi$, $\epsilon=\phi^2$ etc.

\section{Review}

In this section we first briefly review the Ising CFT, focusing on previous results relevant to our study. We then present a model for level repulsion in the context of CFTs.

\subsection{The Ising CFT}

In integer dimensions, the Ising CFT can be defined in terms of the lattice formulation of the Ising model. 
This lattice spin model is described by the following Hamiltonian,
\begin{equation}
\label{H_Ising}
H=J\sum_{\langle ij\rangle }  \sigma_i \sigma_j,
\end{equation}
where $\langle ij\rangle $ indicates that the sum runs over all nearest-neighbor pairs of lattice points and $\sigma_i \in \{-1,+1\}$. We also take $J<0$ so that the energy is minimized when the spins align.

At the continuous phase transition that is present in this model all correlation functions become scale invariant at distances much larger than the lattice spacing. In the continuum limit, these correlation functions are believed to be described by a CFT.\footnote{In the special case of $d=2$ this has been proven \cite{Chelkak2012}. For developments in $d=3$ and $d=4$ see \cite{Aizenman2014} and \cite{Aizenman:2019yuo} respectively.} As can be seen from \eqref{H_Ising} this theory possesses a $\mathbb{Z}_2$ flavor symmetry acting on the spin degrees of freedom.

Taking the continuum limit one obtains a system described by the following Hamiltonian density (see for instance \cite{Fisher1983})
\begin{equation}
\mathcal{H} = ( \partial \phi )^2 +m^2 \phi^2 +g \phi^4 \, .
\label{hamiltonian}
\end{equation}
This quantum field theory can be studied in $d=4-\varepsilon$ dimensions by analytic continuation. In the limit $\varepsilon \rightarrow 0$, this theory possesses an RG fixed point in the IR, at which $m^2=0$ and $g=O(\varepsilon)$. The fixed point is of course scale invariant, since the beta function vanishes there. Thus, under the assumption that the scale invariance is enhanced to conformal invariance, this fixed point should again be described by a $\mathbb{Z}_2$ symmetric CFT. 
Since this QFT describes the continuum limit of the Ising model, its IR fixed-point is the Ising CFT. 
Working at small $\varepsilon$ the coupling constant at the fixed-point is infinitesimally small and provides the basis for a controlled perturbative expansion.

An alternative to this perturbative understanding of the Ising CFT, is the more axiomatic approach of the conformal bootstrap. In this approach conformal symmetry is imposed and the Ising CFT is obtained as the only unitary interacting CFT with $\mathbb{Z}_2$ symmetry that has $\Delta_\sigma$  and $\Delta_\epsilon$ values close to those expected from other descriptions of the Ising CFT.\footnote{This last condition is required because the numerical conformal bootstrap cannot exclude solutions at large values of $\Delta_\sigma$ where an additional peninsula always exists. An example of a theory populating this peninsula is the supersymmetric Ising model \cite{Fei:2016sgs,Rong:2018okz,Atanasov:2018kqw, Atanasov:2022bpi}. 
}

\subsubsection{The $\eps$-expansion}
In this work, we will be comparing with various results obtained using the $\varepsilon$-expansion. The techniques used to obtain these can roughly be placed in three groups. 

The first is the usual Feynman diagram expansion. This method provides the data to highest order in $\varepsilon$. 
Calculations up to six loops for generic multi-scalar theories have been carried out in \cite{Bednyakov:2021ojn}, building on the calculations of \cite{Kompaniets:2017yct}. Seven loop results specifically for $\Delta_\phi $, $\Delta_{\phi^2}$ and $\Delta_{\phi^4}$ can be found in \cite{Schnetz:2016fhy}. 
The scaling dimensions for $\phi$, $\phi^2$ and $\phi^4$ that are used in this work were obtained with these methods.\footnote{A result for $\Delta_\phi$ has recently been obtained to order $\varepsilon^8$ \cite{SchnetzUnp}.}

The second group consists of calculations that work systematically for generic operators (i.e. generic spin and global symmetry representation) at one loop \cite{Kehrein:1992fn,Kehrein:1994ff}. 
The fact that one can systematically compute corrections to anomalous dimensions of any operator at one loop, without computing explicit diagrams, can be explained by the observation that the beta function at one loop may be completely determined via knowledge of the OPE coefficients at zeroth order \cite{Cardy:1996xt}. 
The free theory OPE coefficients themselves are computable via Wick contractions, see \cite{Hogervorst:2015akt} for an application of this method to the one-loop dilatation problem. 
The result of such methods are lists of primary operators constructed out of powers of the order parameter field $\phi$, with the insertion of gradients, together with results for their one-loop anomalous dimension \cite{Kehrein:1994ff,Henriksson:2022rnm}.

The third method combines the $\varepsilon $ expansion with the analytic bootstrap, which typically gives the CFT data for certain families of operators, referred to as twist families. 
The most precise results concern the leading twist operators. A general recipe for computing CFT data in $\phi^4$ theories with arbitrary global symmetries using this approach has been presented in \cite{Henriksson:2020fqi} (also for large $N$). 
While less systematic than directly computing diagrams, the method has various advantages. For example, one obtains simultaneously the anomalous dimensions for an entire family of operators, i.e. for arbitrary spins with twist and global symmetry representation fixed. 
Another advantage is that the computation of OPE coefficients is often not much more complicated than the computation of scaling dimensions.\footnote{See for instance \cite{Gopakumar:2016cpb,Alday:2017zzv,Carmi:2020ekr}.}

\subsubsection{The Ising CFT in 3d}

One way of obtaining quantitative predictions for conformal data in $d=3$ is to continue the $\varepsilon$-expansion data to $\varepsilon =1$. Typically, in order to obtain somewhat reliable estimates for exponents, one has to perform resummations. 
These estimates are a posteriori found to be in good agreement with other non-perturbative methods such as the conformal bootstrap (see e.g.\ figure~\ref{fig:even0-estimates}).
While this is commonly believed to be the case for leading operator dimensions, in this work we find that, remarkably, this also holds for many operators higher up in the spectrum, even when we only have access to the first-order anomalous dimensions in the $\eps$-expansion.

On the conformal bootstrap side, state of the art results include \cite{Kos:2016ysd}, which confined $\Delta_\sigma $, $\Delta_{\epsilon}$ and $\theta=\arctan(\lambda_{\epsilon\epsilon\epsilon}/ \lambda_{\sigma\sigma\epsilon})$ to lie in a small isolated three dimensional island. These islands were so small, that they provided the most precise determination of the critical exponents describing the Ising critical point. In \cite{Simmons-Duffin:2016wlq}, using the extremal functional method of \cite{El-Showk:2012vjm,El-Showk:2016mxr}, the author was able to estimate the operator data for many low dimensional operators. A large part of these operators were observed to fall on so called double twist trajectories, as predicted analytically in \cite{Komargodski:2012ek,Fitzpatrick:2012yx}.\footnote{In perturbation theory, such operators were observed to exist much earlier \cite{Parisi:1973xn,Callan:1973pu,Kehrein:1995ia,Derkachov:1996ph}.} More recently, \cite{Reehorst:2021hmp} provided rigorous error bars for the determination of some of the operators studied in \cite{Simmons-Duffin:2016wlq}.

\subsubsection{Bootstrapping in non-integer dimensions}

An important idea that we test in this paper is that of  ``spectrum continuity,'' i.e. that the spectrum, as well as other CFT-data such as OPE coefficients, should change continuously as we follow the fixed-point across spacetime dimensions $d$. While success of the $\varepsilon$-expansion for observables like the critical exponents suggests that such continuation is possible, there are some potential caveats. The most important in the context of this paper is the fact that by a careful study of the $\varepsilon$-expansion, the Ising model was found to be non-unitary in non-integer dimensions \cite{Hogervorst:2015akt}. This is manifest in several ways, such as negative-norm states, complex OPE coefficients and complex anomalous dimensions. 
The continuation of the spectrum across dimensions also raises questions about representation theory of the Lorentz (rotation) group $SO(d)$ for non-integer $d$. Since the theory also respects spacetime parity, we can combine Lorentz and parity to $O(d)$, and the representation theory for that group can formally be continued across $d$, see the discussion in \cite{Binder:2019zqc} for the $O(N)$ global symmetry group.\footnote{For $O(N)$ models, continuing in $N$ is in this respect similar to continuing in $d$. Alternatively, the $O(N)$ model at non-integer $N$ can be seen as a loop gas model. See for instance \cite{Grans-Samuelsson:2021uor,Gorbenko:2020xya} for representation theory of the 2d $O(N)$ model, and \cite{Liu:2012ca} for a Monte Carlo study of the 3d $O(N)$ model.}

Despite the violation of unitarity just discussed, in any practical application so far the conformal bootstrap has been insensitive to it. For example, \cite{El-Showk:2013nia} and \cite{Cappelli:2018vir} found that the kinks in exclusion plots persist in fractional dimensions and moreover, the positions of these kinks are in excellent agreement with perturbative predictions \cite{LeGuillou:1987ph}.
Thus, at the derivative orders currently being studied, the bootstrap appears to be essentially insensitive to any non-unitarity of this type. In other words, the numerical conformal bootstrap is unable to exclude these theories from parameter space even though they are non-unitary. The same effect was found to persist also when a mixed correlator setup is used to find isolated islands \cite{Behan:2016dtz}.\footnote{Below $d=2$, on the other hand, the picture changes drastically, and the sequence of clear kinks for $d\geqslant 2$ degenerates into two features that then move away from each other \cite{Golden:2014oqa}.}

A related observation was made in the study of a vector model with $O(500)$ symmetry in $ d=5$ \cite{Li:2016wdp}. This theory is also expected to be non-unitary and yet imposing unitarity an isolated island was found. Note that this non-unitarity is of a different type. Here the non-unitarity comes from large-$N$ instanton corrections to the scaling dimensions, i.e. contributions proportional to $e^{-\frac{a}{1/N}}$ which multiply a complex number, where $a$ is some $d$ dependent number \cite{Giombi:2019upv}. Also in this case, the bootstrap was insensitive to the non-unitarity.

Lastly, a more recent example is \cite{Sirois:2022vth}. In this case the author studied the $O(N)$ model for various non-integer values of $N$, which are know to be non-unitary \cite{Binder:2019zqc}. However, the bootstrap was apparently yet again insensitive to the violation of unitarity in the range $N>1$.\footnote{For $N<1$, there is a clear manifestation of unitary violation in terms of negative squared OPE coefficients involving low-lying operators even in the free theory. In this region, indeed the bootstrap approach of \cite{Sirois:2022vth} was able to detect this unitarity violation.} This trend continues in the present work, where also do not see any noticeable effect due to the violation of unitarity.

\subsection{Level repulsion}
\label{sec:repulsion}

In this section we will give some more details on level repulsion or avoided crossing, which is a generic feature of Hermitian matrices, discussed in the context of quantum mechanics by von Neumann and Wigner \cite{vonNeumann1929}. 
In the context of conformal field theories, the closest equivalent to the Hamiltonian of quantum mechanics is the dilatation operator, and the spectrum of energy levels corresponds to the spectrum of scaling dimensions.\footnote{Level repulsion does not happen when states belong to different symmetry irreps. In what follows, we assume that all operators discussed are primaries in the same representation of the Lorentz and global symmetry group.}
Level repulsion would be manifest as observing the scaling dimensions of two operators, as we vary $\eps$, to first approach each other and then repel without crossing. 

As mentioned in the introduction, level repulsion in the contexts of CFT was discussed in \cite{Korchemsky:2015cyx} with examples from $\mathcal N=4$ SYM, extremal spectra at the bootstrap bound and QCD baryon distributions.\footnote{Moreover, in \cite{Behan:2017mwi}, it was shown that on a one-dimensional conformal manifold, i.e. in the context of an exactly marginal coupling, avoided crossing can be inferred from the differential equation that governs the evolution of the CFT-data.}

Here we will make a simplified argument for how a matrix like \eqref{eq:mixinggenIntro} would arise to a reasonable approximation in a general setup.
Although level repulsion happens at finite coupling -- for us manifested as finite $\eps$ -- it is instructive to start by considering the situation in perturbation theory. 
For concreteness, consider the case of perturbing the free theory with a collection of operators, 
\begin{equation}
\label{eq:H0}
\mathcal H=\mathcal H_0+\sum_i g_i\O_i
\end{equation}
and focus on the case where there is a non-trivial conformal fixed-point, $g_i=g^*_i$. Here, in principle, the sum runs over all operators in the theory. Moreover, we assume that the couplings depend on a single underlying parameter $\eps$, so that $g^*_i\to0$ as $\eps\to0$.\footnote{For us, $\eps$ will be the dimensional parameter $\eps=4-d$, but for instance in the existence of a conformal manifold, $\eps$ could be the coordinate along a one-parameter curve inside the conformal manifold.}
The perturbative spectrum is given by the eigenvalues of the matrix of partial derivatives of the beta functions
\begin{equation}
  \omega=\left.\left(\frac{\partial\beta_i}{\partial g_j}\right)\right|_{g^*}.
\end{equation}
More precisely, 
\begin{equation}
\Delta_i= d-\lambda_i, \qquad   \omega \vec v_i=\lambda_i\vec v_i
\end{equation}
In perturbation theory, typically $  \omega$ is an upper block-triagonal matrix,\footnote{See e.g.~\cite{Brown:1979pq}.} where the blocks are formed out of operators with the same dimension in the free theory,
\begin{equation}
\label{eq:omegablock}
  \omega =
\begin{pmatrix}
\cline{1-1}
   \multicolumn1{|c}{\omega_{n_1\times n_1}} & \multicolumn1{|c}{    A} &      A' \\
\cline{1-2}
&\multicolumn1{|c}{  \omega_{n_2\times n_2} }  &  \multicolumn1{|c}{  A''}&    \ddots
\\
\cline{2-3}
	0&   & \multicolumn1{|c}{  \omega_{n_3\times n_3} }  &  \multicolumn1{|c}{} &  \\
\cline{3-4}
&  &  &   \multicolumn1{|c}{\ddots} 
\end{pmatrix}.
\end{equation}
Here $A$, $A'$ etc.\ denote matrices with entries that are generically non-zero.
The zeros in this matrix may get non-perturbative corrections.\footnote{Such contributions may for instance be generated by instanton contributions of the form $e^{-a/\varepsilon}$, where $a$ is a number independent $\varepsilon$. }
Neglecting such non-perturbative corrections, the set of eigenvalues of \eqref{eq:omegablock} is the collection of the eigenvalues of each of the diagonal blocks. 
The precise form of $ \omega$ is scheme dependent, however its eigenvalues (at the critical point) are not.

To a reasonable approximation, valid \emph{a priori} only at infinitesimal $\eps$, the spectrum of the theory can then be written as the eigenvalues of a matrix
\begin{equation}
\label{eq:bigmixing}
\mathbb D=\begin{pmatrix}
\cline{1-3}
\Delta_1(\eps) & & 0&  \multicolumn1{|c}{} & & &   \\
& \ddots & &  \multicolumn1{|c}{}&     X & &      X'
\\
0& & \Delta_{n_1}(\eps) &  \multicolumn1{|c}{} &&&  \\
\cline{1-6}
& & &  \multicolumn1{|c}{\Delta_{n_1+1}(\eps)} &  & 0&    \multicolumn1{|c}{}
\\
&    X^{\dagger} &  &  \multicolumn1{|c}{} & \ddots & &  \multicolumn1{|c}{  X''}
\\ & & &  \multicolumn1{|c}{0}& & \Delta_{n_1+n_2}(\eps) & \multicolumn1{|c}{}
\\
\cline{4-7}
&    X'^\dagger & & &     X''^\dagger & &   \multicolumn1{|c}\ddots
\end{pmatrix}.
\end{equation}
Here the diagonal elements are the eigenvalues of \eqref{eq:omegablock}, computed block by block, and we have accounted for some off-block-diagonal entries in the form of matrices $  X$, $  X'$, etc, which must vanish as $\eps\to0$.\footnote{It may be that non-perturbative corrections also appear at the ``zero entries'' inside the blocks of \eqref{eq:bigmixing}. 
In that case one can either perform a further change of basis to accommodate them, or reduce the size of the blocks. The conclusions for level repulsion below remain unchanged.} They may get contributions at higher order in perturbation theory, or be completely non-perturbative. 

As long as the off-diagonal terms $X$, $X'$ etc. are small, the set of eigenvalues of $\mathbb D$ in \eqref{eq:bigmixing} is captured well by the diagonal entries, which we take without loss of generality to be ordered by their values at $\eps=0$, $\Delta_i(0)\leqslant \Delta_{i+1}(0)$ etc. 
This statement breaks down when two consecutive diagonal entries in different blocks become approximately equal, which is the case where there is an apparent crossing, say $\Delta_{n_1}(\eps)\approx \Delta_{n_1+1}(\eps)$. In the region where
\begin{equation}
|\Delta_{n_1}(\eps)- \Delta_{n_1+1}(\eps)|\lesssim x
\end{equation}
we can no longer ignore the corresponding off-diagonal entries $x=X_{n_1,1}$ and $X_{n_1,1}^*$. Notice that by construction, this can only happen at finite $\eps$.

As long as no third operator dimension is also in the vicinity, we can simplify the analysis further by looking at the eigenvalues of the following submatrix,
\begin{equation}
\label{eq:mixinggen}
\begin{pmatrix}
\Delta_1(\eps) & x(\eps)\\x(\eps)& \Delta_2(\eps)
\end{pmatrix}.
\end{equation}
This is the matrix introduced in \eqref{eq:mixinggenIntro}.
For simplicity we consider the non-diagonal terms to be real.
The eigenvalues of \eqref{eq:mixinggen} are
\begin{equation}
\label{eq:mixingdims}
\Delta_\pm=\frac{\Delta_1+\Delta_2}2\pm\sqrt{\left(\tfrac{\Delta_1-\Delta_2}2\right)^2+x^2}
\end{equation}
and the minimal distance, where $\Delta_1= \Delta_2$, is 
\begin{equation}
(\Delta_+-\Delta_-)_{\mathrm{min}}=2x  .
\end{equation}

Let us try to convert the preceding schematic discussion into to a testable hypothesis.
Our model \eqref{eq:bigmixing} gives a good description of the spectrum of the theory if the following two conditions hold:
\begin{enumerate}
\item The off-block-diagonal terms in $  X$, $  X'$ etc. are numerically small, say $ \lesssim x \ll 1$.
\item Away from the regions where $|\Delta_{i}(\eps)- \Delta_{j}(\eps)|\lesssim x$, the diagonal entries $\Delta_i(\eps)$, computed from perturbation theory, approximate the spectrum well.
\end{enumerate}
A theory in which these conditions hold can be said to be well approximated by perturbation theory.\footnote{It may also happen that $x$ is numerically small, but the perturbative predictions of $\Delta_{i}(\eps)$ fail to reproduce the spectrum at finite $\eps$. 
In that case we would still expect level repulsion to occur, and expect the two operators involved in the repulsion to have dimensions locally described by the eigenvalues \eqref{eq:mixingdims}, as long as $\Delta_i$ and $x$ do not change too rapidly in $\eps$.}
In fact, a case where the discussion above can be put on a more firm footing is in $\mathcal N=4$ SYM, at large number of colors $N$ \cite{Korchemsky:2015cyx}. 
There the diagonal elements of \eqref{eq:bigmixing} are not computed from perturbation theory, but can instead be taken to be the planar operator dimensions, known from integrability \cite{Gromov:2009zb}. In that case, the off-diagonal terms scale as $O(1/N)$, leading for instance to level repulsion at order $1/N$ between the Konishi operator and a sequence of double-trace operators with increasing scaling dimension. In fact, by using holographic results for OPE coefficients from \cite{Minahan:2014usa}, the minimal separation in the case of mixing between Konishi and the singlet double-trace operator $[\O_J,\O_J]$ at large $J$ was determined to leading order in $1/N$ \cite{Korchemsky:2015cyx}.

A concrete question that we ask in this paper is if we can numerically observe the phenomenon of level repulsion, and then evaluate to what extent the two above conditions hold. 
We stress that it is not obvious that they should apply to our case. 
In fact, since strictly speaking we are considering a different theory at each different dimensions $d$, we cannot exclude the possibility of other, more exotic, scenarios to occur, such as operators annihilating and going into the complex plane, or decoupling from the spectrum. Let us point out that, to some extent, both these things do indeed happen in the Ising CFT; there are operators with complex scaling dimensions, and operators that vanish identically when one approaches integer spacetime dimensions \cite{Hogervorst:2015akt}. 
These effects, however, are expected to happen higher up in the spectrum than what we are able to probe in this work.

\section{Method}

For the rest of this work we will make use of standardised numerical bootstrap techniques, which will be briefly outlined below as well as in the appendices. 
We study the same crossing equations as \cite{El-Showk:2012cjh},\footnote{These may also be obtained in an automated fashion with \cite{Go:2019lke}, which also works for a large amount of supported groups beyond $\mathbb{Z}_2$.} but instead of studying only the case $d=3$ we will consider fractional dimensions between $2.6$ and $4$. Such a comprehensive multi-correlator study in such a wide region of dimensions has not been done before and is now possible due to advancing numerical bootstrap techniques. Specifically we will be relying on a a recent advancement in the numerical conformal bootstrap called the \emph{navigator} \cite{Reehorst:2021ykw}.

\subsection{Obtaining the spectrum using the navigator and the EFM}

The navigator, or navigator function, is a function that replaces the binary `allowed' /`disallowed' information from the traditional bootstrap with some continuous measure of success. I.e., it gives some measure of the distance to an allowed point. Specifically, for the search space spanned by the dimension of the relevant $\mathbb{Z}_2$-odd and $\mathbb{Z}_2$-even scalars $\Delta_\sigma$ and $\Delta_\epsilon$ and the OPE angle $\theta=\arctan(\lambda_{\epsilon\epsilon\epsilon}/ \lambda_{\sigma \sigma \epsilon })$, \cite{Reehorst:2021ykw} constructed a function $\mathcal{N}(\Delta_\sigma,\Delta_\epsilon,\theta)$, with the following properties:
\begin{enumerate}
	\item The function is bounded from above and exists in the whole search space.
	\item It differentiates between disallowed and allowed points by being positive on the former and non-positive on the latter. (Being zero on the boundary by continuity.)
	\item It is differentiable.
	\item It is cheaply computable by standard technology (it corresponds to an OPE maximization using SDPB \cite{Simmons-Duffin:2015qma,Landry:2019qug})
	\item Its derivatives are cheaply computable (see \texttt{approx\_objective} provided with SDPB).
\end{enumerate}
These properties allow us to efficiently find isolated allowed regions (i.e. islands) and the boundaries of these regions. It also allows us to find the minimum of the navigator function which was conjectured to lie close in the search space to the true underlying CFT.\footnote{This can be seen for example by considering how the island and the navigator function change when changing the derivative order used in the numerics.} Here we conjecture this to also extend to the extremal spectrum. Specifically, 
we conjecture that the solution to the (truncated) crossing equations, extracted by the extremal functional method (EFM), at this minimum point, is a good predictor for the spectrum of the underlying CFT.\footnote{In the numerical conformal bootstrap a functional is constructed from a certain number of derivatives at $z=\bar{z}=\frac{1}{2}$. A parameter $\Lambda$ is usually used to parameterize the number of derivatives in this basis of functionals. A larger value of $\Lambda$ leads to stronger bounds. Bounds derived at any value of $\Lambda$ are rigorous and will hold also for the solution to the full crossing equations. On the other hand a solution extracted by the EFM method will only be a solution to the truncated crossing equations given by the derivative expansion around the crossing-symmetric point. This solution is expected to give a good approximation to the low-dimensional operator spectrum of the solution to the full crossing equations.} Thus, we use the following procedure in order to estimate the spectrum at all values of $\varepsilon$:
\begin{enumerate}
	\item Minimize the navigator function $\mathcal{N}(\Delta_\sigma,\Delta_\epsilon,\theta)$. (Using the modified Broyden--Fletcher--Goldfarb--Shanno (BFGS) algorithm proposed in \cite{Reehorst:2021ykw}).
	\item Assuming a negative minimum is found, minimize the central charge $C_T$ in order to obtain an extremal functional.
	\item Extract this extremal solution to the truncated crossing equations using for example \texttt{spectrum.py} \cite{spectrum}.
\end{enumerate}
We choose to consistently present only the spectrum at the navigator minimum rather than searching for many allowed points to minimize $C_T$ at and averaging over the resulting spectra. The navigator minimum provides a clear universal criterion to pick a point. 
Since the spread in the data found of spectra obtained at different points is small, this should not qualitatively affect our study. 
Given that the navigator point is the furthest from being excluded at that given derivative order, we expect it to be a better predictor of the true spectrum than any other point. We leave it for future work to apply the methods of \cite{Reehorst:2021hmp} in order to turn our results into rigorous bounds.

When constructing the navigator function we use the following assumptions:
\begin{align}
\Delta_{\epsilon'} &\geqslant d
\label{assumption1}\,,\\
\Delta_{\sigma'} &\geqslant d
\label{assumption2}\,, \\
\Delta_{T} &= d \,,\\
\Delta_{T'} &\geqslant d+1\,.
\label{assumption3}
\end{align}
Here $\Delta_{\epsilon'}$ denotes the dimension of the second lowest $\mathbb{Z}_2$-even scalar, $\Delta_{\sigma'}$ the second lowest $\mathbb{Z}_2$-odd scalar and $\Delta_{T'}$ the dimension of the second lowest spin-2 $\mathbb{Z}_2$-even operator. We also assume the existence of the stress tensor $T$. Finally we impose the stress-tensor ward identity, i.e. we impose that the ratio of OPE coefficients with which the stress tensor appears in the $\sigma \times \sigma$ and $\epsilon \times \epsilon $ OPEs is fixed as follows \cite{Osborn:1993cr}:
\begin{equation}
\lambda_{\sigma \sigma T} / \lambda_{\epsilon \epsilon T} =\Delta_{\sigma} / \Delta_{\epsilon}.
\label{assumption4}
\end{equation}
For the case of $d=3$ these assumptions are identical to the assumptions made in \cite{Simmons-Duffin:2016wlq}.

For more details on how these assumptions lead to the construction of the navigator function $\mathcal{N}(\Delta_\sigma,\Delta_\epsilon,\theta)$ see appendix~\ref{app:nav}.

\subsection{Initial search point}
Given that we only wish to reach the minimum of the navigator function, and not map out in detail the parameter space in the vicinity of the Ising CFT, we started the navigator search as close as possible to the actual CFT. This reduces the number of navigator function calls required to find the minimum and increases numerically efficiency. To this end we used all previously available information to make an optimal guess for the position of the allowed island in various dimensions. In practice, we performed a Pad\'e resummation of the $\varepsilon^7$ results of \cite{Schnetz:2016fhy} such that in $d=3$ they match the results of \cite{Kos:2016ysd}.\footnote{For the sake of clarity, we stress that in Pad\'e resummations throughout the rest of the work we do \emph{not} fix the behavior at $d=3$.} We then took the average between these, and the values given by the interpolation of \cite{Cappelli:2018vir}. Similarly, for the OPE coefficients we used the results of \cite{Carmi:2020ekr} and \cite{Henriksson:2020jwk}, and then required that the resummation matched the 3d value known from the bootstrap.

\subsection{Extremal functional method}
When we use semi-definite programming to find a lower bound on $C_T$ at the navigator minimum this provides an extremal functional \cite{El-Showk:2012vjm,El-Showk:2016mxr,Simmons-Duffin:2016wlq}. 
The zeros of the determinant of this functional acting on the $\vec{V}$'s appearing in \eqref{Vepstheta} should correspond to the dimensions of the operators in the solution to the (truncated) crossing equations at this point. We extract this spectrum using standard procedures (see \cite{spectrum}). 
Due to finite numerical errors in the solution, the positions of the double zeros of the functional will not really reach zero. Therefore it is customary to look for the minima of the determinant instead. 
One caveat should be noted: \cite{spectrum} excludes operators whose squared OPE coefficients are negative. However, we found that there was often a numerical instability due to which the OPE coefficients could not accurately be determined from the primal solution. 
Therefore, the OPE coefficients of physical operators could appear to be negative and wrongly be excluded. In order to not miss any operators we instead report all minima whether \texttt{spectrum.py} finds a OPE coefficient squared that is positive or not, see also Appendix~\ref{sec:EFM}. 
Note that in order to obtain an estimate of the OPE coefficients this method minimizes the error in an over determined system. In principle all the equations in this overdetermined system should approximately agree on the correct OPE coefficients. However, we found that for some points this was not the case. Picking some subset of equations and minimizing the error only for that subset results in a large spread in the OPE coefficients found. The OPE coefficient can easily change many magnitudes and even the sign depending on which equations in the overdetermined system are included in the error minimization.\footnote{This phenomenon also occurs in $d=3$ and we therefore believe it to be unrelated to actual negative squared OPEs appearing in the solution.}

\section{Results}

In this section we present the results from the numerical study of this paper, and compare with perturbative data. For all $d$, a single negative navigator minimum was found in the three-dimensional space parametrized by ($\Delta_\sigma$, $\Delta_\epsilon$, $\theta$). We present these in the section below. Next we study the spectra extracted from these points, starting with the $\mathbb Z_2$-even scalars where a clear example of level repulsion is observed. We then briefly comment on the spectra in some other interesting channels. Finally, we look at data for the leading two OPE coefficients.

\subsection{Navigator minima} 

The first step in the determination of the spectrum is to find a set of allowed points. We performed this study at dimensions 
\begin{equation}
d\in\{2.6,\,2.7,\,2.8,\,2.9,\,3.0,\,3.1,\,3.2,\,3.3,\,3.4,\,3.5,\,3.6,\,3.7,\,3.8,\,3.9,\,3.95\}.
\end{equation}
At each of the points we find a local navigator minimum where the navigator is negative. This gives a sequence of points
\begin{equation}
\label{eq:navigatorMinimaSchematic}
\left(\Delta_{\sigma}|_{\mathrm{min}}(d),\Delta_{\epsilon}|_{\mathrm{min}}(d),\theta|_{\mathrm{min}}(d)\right),
\end{equation}
where $\theta=\arctan(\lambda_{\epsilon\epsilon\epsilon}/ \lambda_{\sigma \sigma \epsilon })$. The complete collection of allowed points are reported in table~\ref{tab:navigatorminima} in appendix~\ref{sec:navigatorminima}. 
In figure \ref{fig:finalpoints} we display the corresponding anomalous dimensions $(\Delta_{\sigma}|_{\mathrm{min}}-\frac{d-2}2,\Delta_{\epsilon}|_{\mathrm{min}}-(d-2))$. This figure can be compared with the results of \cite{El-Showk:2013nia}, who gave a corresponding figure with kinks in the $\Delta_{\sigma},\Delta_{\epsilon}$ exclusion plots.
\begin{figure}
\centering
\includegraphics[width=0.8\textwidth]{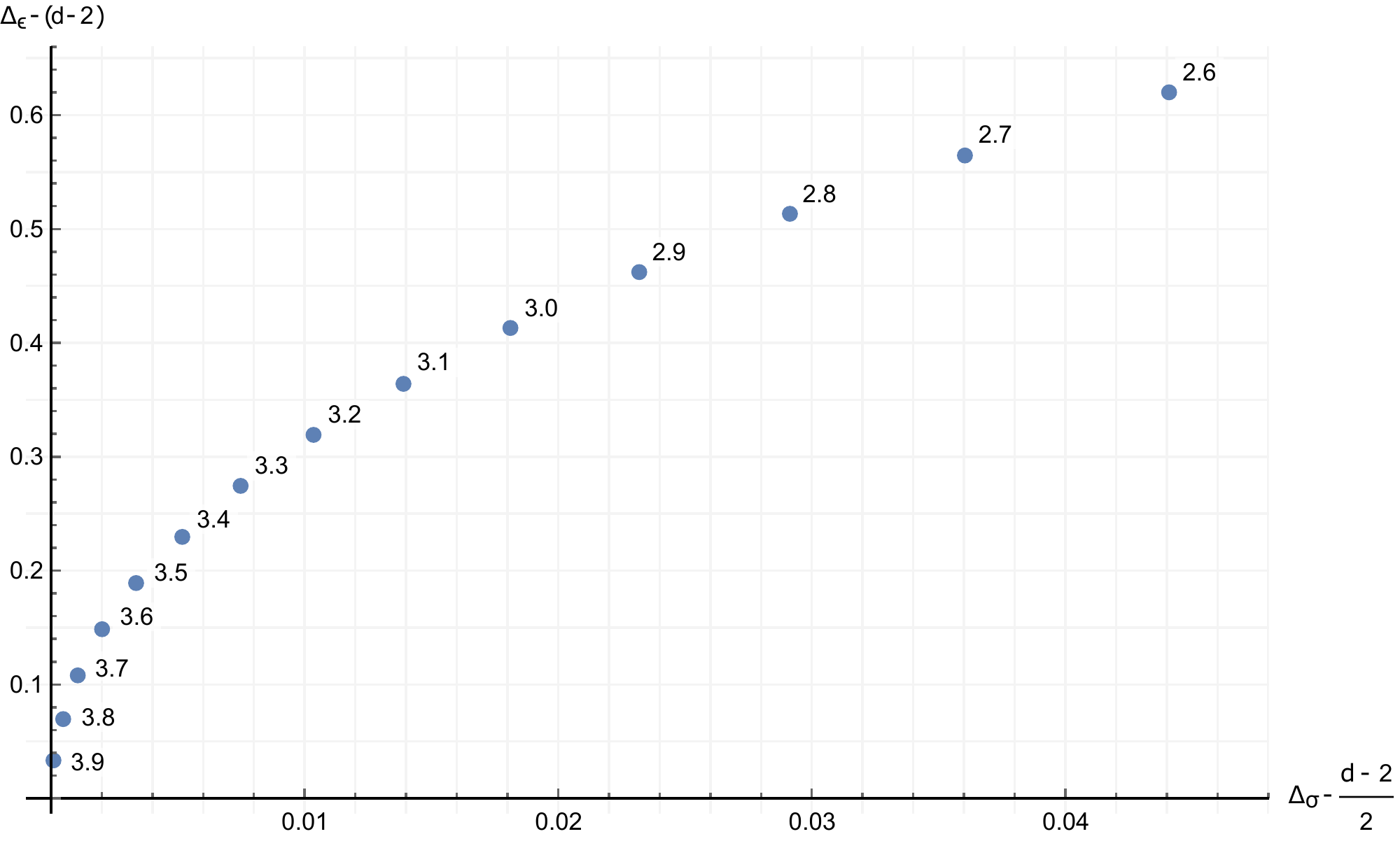}
\caption{The final points of the navigator search, presented in the space of anomalous dimensions. The point $d=3.95$ is excluded for clarity.}\label{fig:finalpoints}
\end{figure}

The points \eqref{eq:navigatorMinimaSchematic} provide estimates for the critical exponents using well-known formulas, see e.g. \cite{Cardy:1996xt,Henriksson:2022rnm}.

\subsection{$\mathbb{Z}_2$-even scalars}

\begin{figure}
\centering
\includegraphics[width=0.95\textwidth]{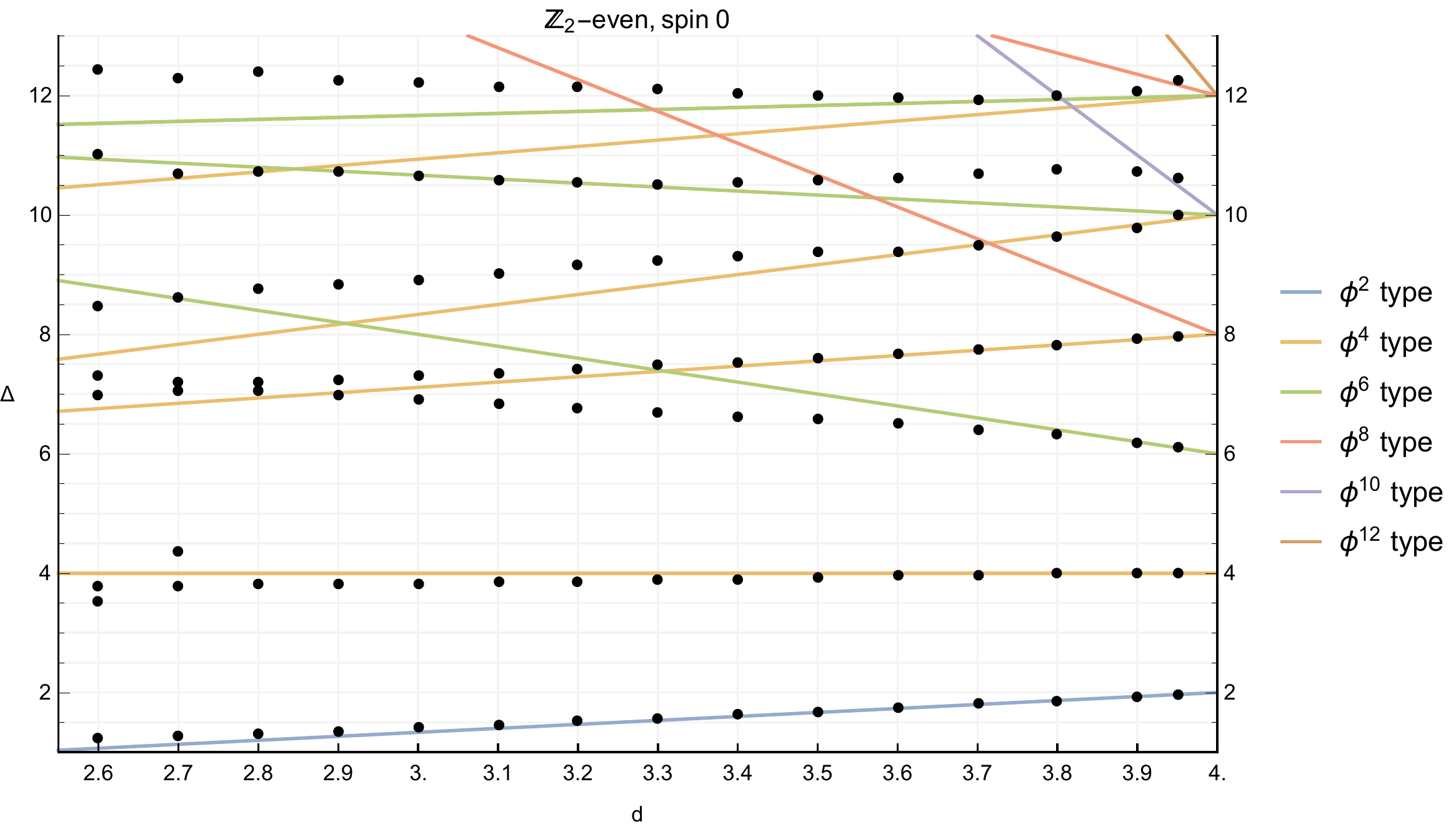}
\caption{The dimensions of the $\mathbb{Z}_2$-even operators of spin 0. The black dots indicate the spectrum found using the EFM at the navigator minimum. The lines of various colors denote the leading order $\varepsilon$-expansion prediction for operators of a certain type (see the main text).}\label{fig:even0}
\end{figure}

At each of the allowed points \eqref{eq:navigatorMinimaSchematic} we determined an estimate for the spectrum using the extremal functional method, as described in appendix~\ref{sec:EFM}. Here we will discuss in detail the spectrum of $\mathbb{Z}_2$-even scalars, which is shown in figure~\ref{fig:even0}.

In this figure, the black dots represent the numerical data corresponding to the minima of the determinant of the extremal functional for $\mathbb{Z}_2$-even scalars in each dimension. The lowest point displayed at each $d$ is the value of the external dimension $\Delta_\epsilon$. 
Figure~\ref{fig:even0} also contains perturbative data for the first twelve $\mathbb{Z}_2$-even scalar conformal primary operators, with their scaling dimensions truncated to order $\eps$. These estimates are shown as  straight lines in the figure. We use color coding to indicate how the operator is constructed in the limit $d\to4$. Specifically, we say that an operator is of ``$\phi^k$ type'' if it is constructed out of $k$ fields $\phi$, with the potential insertion of partial derivatives. The number of partial derivatives may be inferred from the dimension of the operator at $d \rightarrow 4$.

\begin{table}
\caption{$\mathbb Z_2$-even scalars. The third column summarizes the known $\varepsilon$-expansion results while the fourth column refers to evaluations of various resummations at $\eps=1$. Specifically, the evaluation of the hypergeometric--Meijer resummation for $O(\eps^7)$ results, $\text{Pad\'e}_{[1,1]}$ for $O(\eps^2)$ result (see \eqref{eq:padegeneral}), and truncation for $O(\eps)$ results. The fifth column contains the dimensions found in this work at $d=3$. The last column contains know bootstrap results at $d=3$ from previous works. The errors in bold refer to rigorous bounds.
}\label{tab:EvenScalars}
\centering
\begin{tabular}{|c|c|lr|lr|ll|ll|}
\hline
 $i$ & $\O$ & \multicolumn{2}{c|}{$\Delta_\O$} & \multicolumn{2}{c|}{$\varepsilon$-expansion} &  & \hspace{-12pt}This work &  \multicolumn{2}{c|}{Bootstrap}
\\\hline
 1 & $\phi^2$ & $2-\frac{2 \eps }{3}
$   & ($\eps^7$)\cite{Schnetz:2016fhy}  &  $1.4121(6)$    &    \cite{Shalaby:2020xvv} &&  $1.41263$ & $1.412625(\pmb{10})$ & \cite{Kos:2016ysd}
\\
 2 & $\phi^4$ & $4-\frac{17 \eps^2}{27}$ & ($\eps^7$)\cite{Schnetz:2016fhy}   & $3.8231(5)$   &   \cite{Shalaby:2020xvv}  && $3.82978$  &  $3.82951(\pmb{61}) $ & \cite{Reehorst:2021hmp}
\\
 3 & $\phi^6$ & $6+2 \eps
$ & ($\eps^2$)\cite{Derkachov:1997gc} & $6.5918$  &  & &  $6.90348$ &$6.8956(43)$& \cite{Simmons-Duffin:2016wlq}
\\
 4 & $\square^2\phi^4$ & $8-\frac{8 \eps }{9}$ & ($\eps^1$)\cite{Kehrein:1994ff}&   $7.1111$   &  &  & $7.29743$ & $7.2535(51)$ & \cite{Simmons-Duffin:2016wlq}
\\
 5 & $\phi^8$ &  $8+\frac{16 \eps }{3} 
$& ($\eps^2$)\cite{Derkachov:1997gc}& $9.4135$ &   &  \multicolumn{1}{c|}{5} & $8.92059$& $ \sim 8.55$& \cite{Simmons-Duffin:2016wlq}
\\
 6 & $\square^3\phi^4$ &  $10-\frac{5 \eps}{3}$&($\eps^1$)\cite{Henriksson:2022rnm}& $8.3333$  & &    \multicolumn{1}{c|}{6}  &$10.65903$& $\sim10.45$ & \cite{Simmons-Duffin:2016wlq}
\\
 7 & $\square^2\phi^6$ &  $10+\frac{2 \eps }{3}$ & ($\eps^1$)\cite{Kehrein:1994ff}& $10.6667$  & &   \multicolumn{1}{c|}{7} & $12.21187$   &   $\sim11.6$& \cite{Simmons-Duffin:2016wlq}
\\
 8 & $\phi^{10}$ & $10+10 \eps 
$ & ($\eps^2$)\cite{Derkachov:1997gc} & $12.3126$  & &    \multicolumn{1}{c|}{8} & $14.91822$
&&
\\\hline 
\end{tabular}
\end{table}

Studying figure~\ref{fig:even0} we can immediately make the following observations:
\begin{itemize}
\item The numerical results capture correctly the existence of the first four operators that match with the perturbative data in the limit $\eps\to0$. The only exceptions are the points $d=2.6$ and $d=2.7$, where the extremal functional contains an additional zero at $\Delta=3.5379$
 and 
$\Delta= 4.3628$
 respectively. These are likely to be spurious (i.e. they would disappear at a higher derivative order). Apart from these extra zeros, the scaling dimensions of the first four operators seem to follow continuous functions of $d$. 
\item The numerical results completely miss the existence of the fifth operator, which in perturbation theory is the operator $\phi^8$. However they do appear to capture some features that may correspond to the next two operators. In section~\ref{sec:missing} below we give a possible explanation for why $\phi^8$ is the first operator to be missed. 
\item The numerical data for the first four operators match well with estimates for the first four operators from perturbation theory, and this agreement persists also when $\eps$ is not small. In figure~\ref{fig:even0-estimates} we compare the numerical spectrum with perturbative estimates. They are done truncating the perturbative series, and by working out Pad\'e approximants of the form
\begin{figure}
\centering
\includegraphics[width=0.95\textwidth]{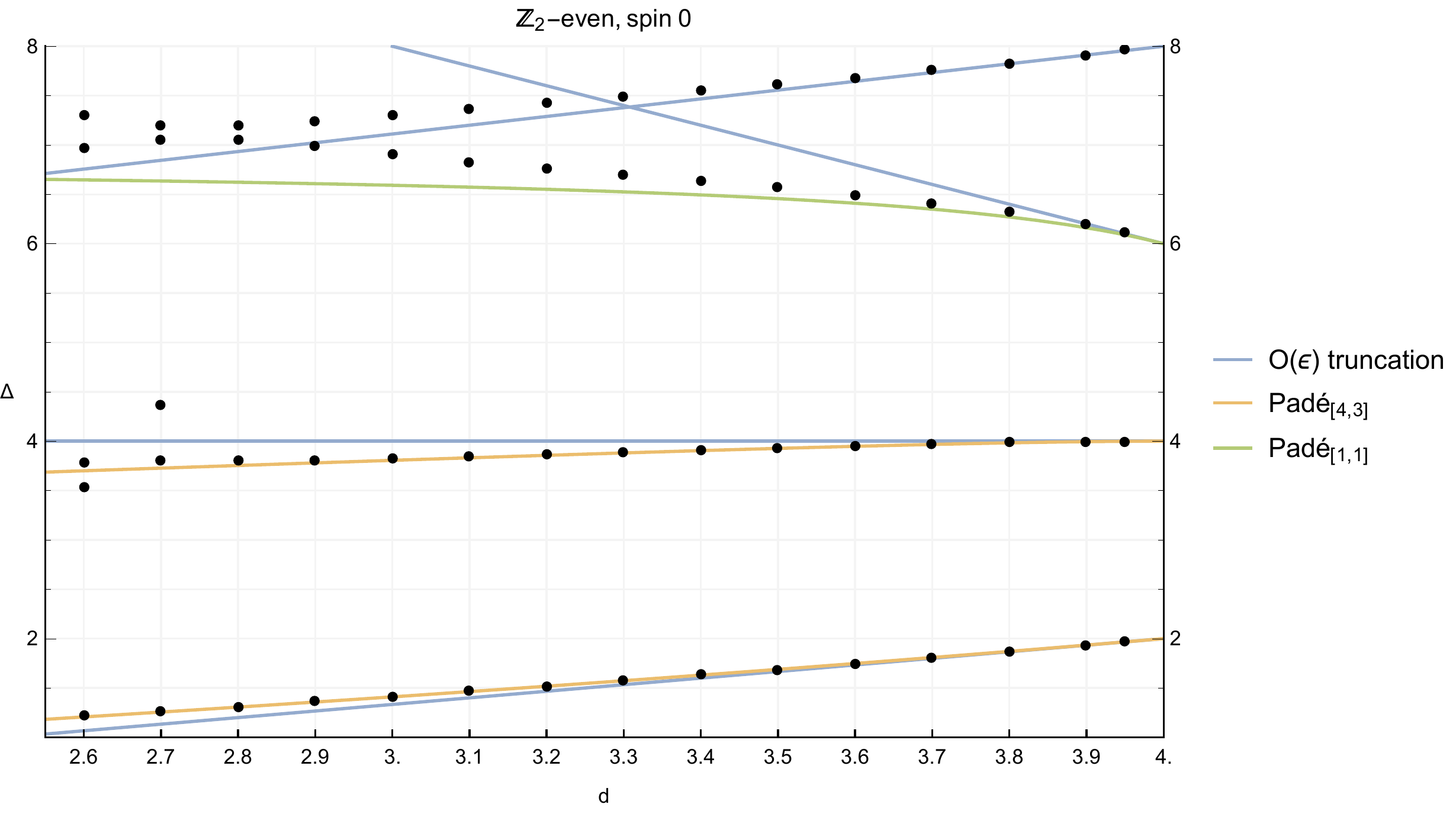}
\caption{Resummed perturbative estimates for the first four $\mathbb Z_2$-even scalar operators. For the fourth lowest operator, there are no results available beyond leading order in $\eps$. The black dots indicate the spectrum found using the EFM at the navigator minimum.}\label{fig:even0-estimates}
\end{figure}
\begin{equation}
\label{eq:padegeneral}
\text{Pad\'e}_{[m,n]}(\eps)=\frac{a_0+a_1\eps+\ldots+a_m\eps^m}{1+b_1\eps+\ldots+b_n\eps^n}.
\end{equation}
A list of perturbative data and estimates is given in table~\ref{table:improvement}. Note that for $\phi^6$ (like for all higher $\phi^k$ operators), both the $O(\eps)$ and $O(\eps^2)$ terms in the anomalous dimensions are numerically large, leading perhaps to somewhat worse predictions.
\end{itemize}

As discussed in appendix~\ref{sec:systematicErr}, the precise numerical values corresponding to the operator dimensions are expected to change if the numerical precision is increased. We believe that the qualitative picture involving the first few operators will not change. We include the raw-data in an ancillary file; more details are given in appendix~\ref{app:rawdata}.

\subsubsection{Level repulsion}
\label{sec:levelRepRes}

Figure~\ref{fig:even0} shows that the scaling dimensions for the third and the fourth operators approach each other and reach a minimum separation in the region $2.6\lesssim d\lesssim 3$. The apparent behaviour of these two sequences of values looks like the repulsion effect that we aimed to investigate in this paper.\footnote{Looking at just the order $\eps$ dimensions corresponding to the straight lines in figure~\ref{fig:even0}, we were initially predicting that we would see this effect above $d=3$. The fact that the strongest mixing happens slightly below $d=3$ led us to extend our results to $d=2.6$.} 

In order to compare with the model \eqref{eq:mixinggen} of a $2\times2$ mixing matrix, we would like to fit a curve that approximates the observed scaling dimensions in the vicinity of the repulsion region. Since for constant $x$ and $\Delta_i$ linear in $\eps$, the eigenvalues of \eqref{eq:mixinggen} define a hyperbola in the $(d,\Delta)$ plane, we fit the numerical data with a conic section. To this end, we search for a $\mathcal Q(d,\Delta)=0$ where
\begin{equation}
\mathcal Q(d,\Delta)=Ad^2+Bd\Delta+C\Delta^2+Dd+E\Delta+F
\end{equation}
We use the fitting procedure described in \cite{Bookstein1979} to perform a least square fit of a conic section:
\begin{equation}
\text{Minimize} \quad \tilde{\chi}^2 \equiv \sum_i\mathcal Q(d_i,\Delta_i)^2\qquad\text{subject to} \quad A^2+\frac {B^2}2+C^2=2
\end{equation}
The sum goes over the operators used for the fit, in our case selected in the range $2.6\leqslant d\leqslant2.9$ and $6<\Delta<8$. A list of these data points is given in appendix~\ref{app:opsforfit}.

\begin{figure}
\centering
\includegraphics[width=0.95\textwidth]{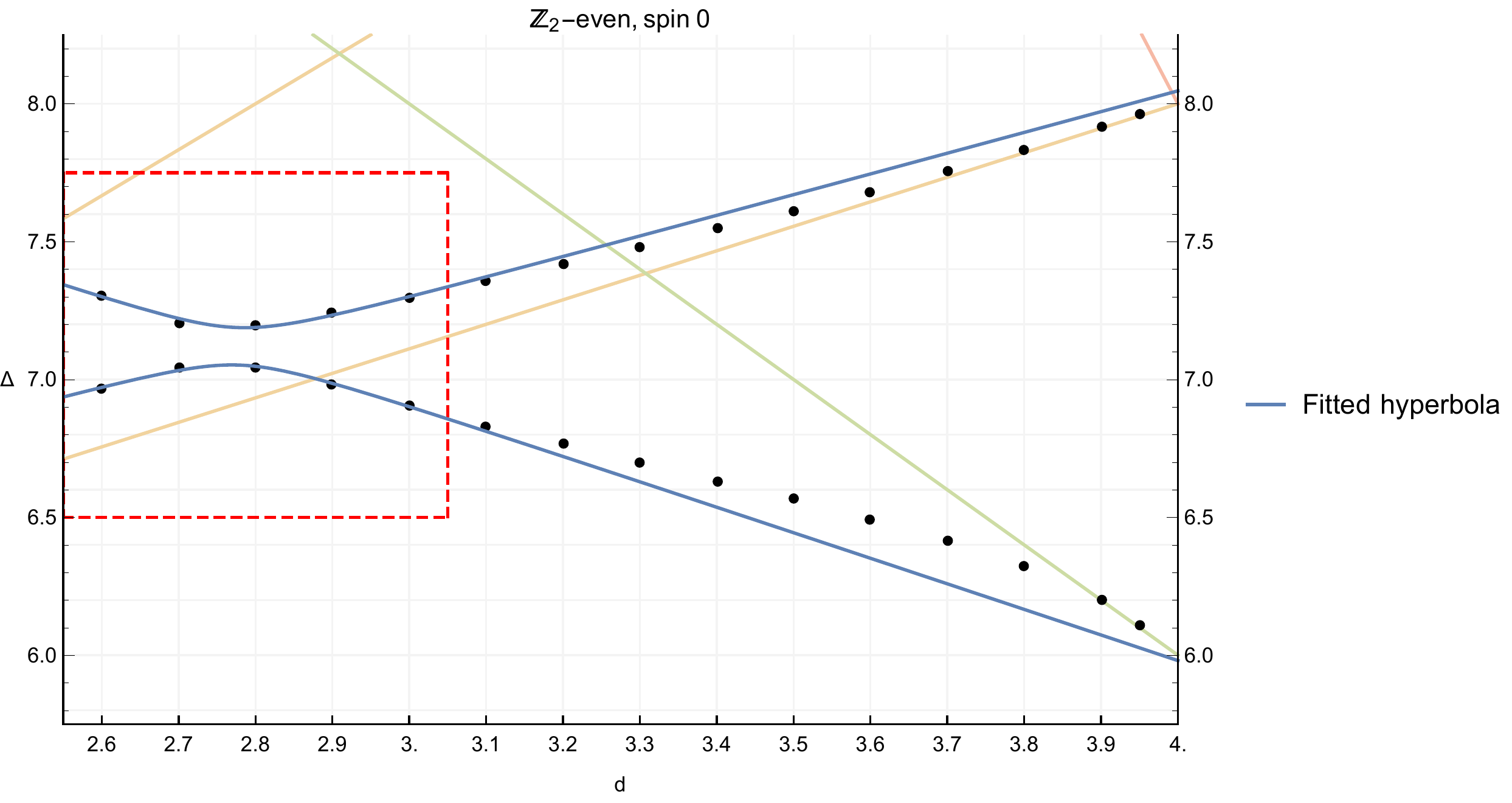}
\caption{Magnified version of figure~\ref{fig:even0}, focusing on the level repulsion effect. The blue curve is given by equation \eqref{eq:fittedHyperbolaEigs} and has been fitted using the operator dimensions in the red box only.}\label{fig:repulsion}
\end{figure}

We find the values
\begin{align}
A&=-0.809771, & B&=0.200861, & C&=-1.150695,
\\
D&=-3.068248,& E&=16.94456, & F&=-56.05891.
\end{align}
The resulting conic section is a hyperbola seen in figure~\ref{fig:repulsion}. 

The minimal difference between the two branches of the fitted curve happens at $d=2.777601$ and takes the value 
\begin{equation}
(\Delta_{\O_4}-\Delta_{\O_3})_{\mathrm{min}}=0.135915.
\end{equation}

To connect with \eqref{eq:mixinggen}, we rewrite $\mathcal Q(d,\Delta)=0$ describing the hyperbola in a more suggestive form, to find
\begin{equation}
\label{eq:fittedHyperbolaEigs}
\det\left(\Delta \1-\begin{pmatrix}
5.982656+0.930688\epsilon  & 0.067957 \\
0.067957 & 8.044623-0.756132\epsilon 
\end{pmatrix}
\right)
=0
\end{equation}
We observe, as seen in figure~\ref{fig:repulsion}, that as $\eps\to0$ the two eigenvalues approach the values $\Delta^{4d}\approx 6$ and $\Delta^{4d}\approx 8$ respectively to quite good precision. This surprisingly good agreement may be somewhat accidental.

It useful at this point to make some remarks. From figure~\ref{fig:even0-estimates} we can draw a few general insights about field theory in general, and the validity of the $\varepsilon$-expansion specifically. We see that the two lowest lying operators, namely $\phi^2$ and $\phi^4$, have their scaling dimensions very well approximated by the perturbative predictions, especially after resummations. Consequently, $\phi^6$ is observed to agree well with the $\varepsilon$-expansion predictions very close to $d \rightarrow 4$ (e.g. for $d=3.95$ and $d=3.9$). Furthermore, this agreement is improved once one performs a resummation, e.g. we have good agreement down to $d=3.6 \sim 3.5$. However, we see that as we approach the mixing region the perturbative prediction becomes progressively worse, as expected, since this mixing was not taken into account when calculating this scaling dimension perturbatively. This teaches a general lesson, one should be careful about taking $\varepsilon$-expansion estimates at face value, at least ones that do not take mixing between finitely separated (in scaling dimension) operators in $d=4$. We stress, though, that this is not necessarily a breakdown of the $\varepsilon$-expansion itself, the agreement could possibly be restored by computing the off-diagonal elements $X$,$X'$,... in \eqref{eq:bigmixing}. The calculation of the mixing curve in figure~\ref{fig:even0-estimates} within the $\varepsilon$-expansion is an interesting open problem.\footnote{Since the Ising CFT is expected to be (slightly) non-unitary at non-integer dimensions one could also consider non-hermitian corrections to \eqref{eq:bigmixing}. This may lead to scaling dimensions that also have an imaginary part, in addition to their real part. Hence, as a function of the dimensionality $d$ we may expect the operator dimensions to move into the complex plane. Such a scenario may be accommodated for by considering a submatrix of the form
	\begin{equation}
	\label{eq:mixinggenAntisymm}
	\begin{pmatrix}
	\Delta_1(\eps) & -x(\eps)\\x(\eps)& \Delta_2(\eps)
	\end{pmatrix},
	\end{equation}
	leading to the following prediction for the scaling dimensions
	\begin{equation}
	\label{eq:mixingdimsImaginary}
	\Delta_\pm=\frac{\Delta_1+\Delta_2}2\pm \sqrt{\left(\tfrac{\Delta_1-\Delta_2}2\right)^2-x^2}\,.
	\end{equation}
The equation above describes an hyperbola with horizontal transverse axis, i.e. a $90^{\circ}$ rotation compared to the best conic-section fit shown in figure \ref{fig:repulsion}, with the dimensions moving of to the complex plane when $\Delta_1$ closely approaches $\Delta_2 $. Our fitting procedure is agnostic to which scenario is correct since it searches the best fit in the space of all conic-sections which also includes these types of hyperbolas as well as the straight lines one would expect in a ``crossing" scenario. The scenario in \eqref{eq:mixingdims} appears to be clearly numerically favored by a much better fit. The fact that scenario leads to a much better fit can easily be seen by eye, but can also be seen numerically by comparing the $\tilde{\chi}^2$ values of $0.000252652$ and $0.000420454$ respectively. Similarly a fit using two straight lines, i.e. a degenerate conic-section, to account for a ``crossing" scenario has an even larger $\tilde{\chi}^2$ of $0.0055818$. These comparisons should come with the qualification that large non-hermitian corrections to \eqref{eq:bigmixing} are essentially already excluded by the assumptions that go into our numerical setup. In the sense that if large non-unitarities existed, we wouldn't find an island in the first place. 
Hence providing evidence that to good approximation the low dimensional operators, to which the bootstrap is most sensitive, are indeed unitary.}

\subsubsection{Missing operators}
\label{sec:missing}

Let us now turn to the topic of operators existing perturbatively in $4-\eps$ dimensions that are not visible in our numerical results. In the $\Z_2$-even scalar sector, the first such operator is $\phi^8$ with dimension
\begin{equation}
\Delta_{\phi^8}=8-4\eps+\frac{28}3\eps-\frac{1198}{81}\eps^2+O(\eps^3)
\end{equation}
determined in \cite{Derkachov:1997gc}. 

Following the picture discussed in section~\ref{sec:EFM}, one may discuss a few general features of the extremal functional predictions. For each sector, i.e. for each irrep of the global symmetry at a given spin, it has been empirically observed that typically only the first few minima\footnote{In terms of increasing scaling dimension.} correspond to actual physical operators, whereas subsequent minima correspond to random/spurious operators. In general it has been observed that the numerical conformal bootstrap becomes progressively less constraining at higher values of the scaling dimensions. This has the following intuitive understanding. The leading operators provide large contributions to the sum rules, while operators higher up in the spectrum provide smaller contributions. As a consequence the numerics do not fix the latter contributions as tightly. 

It is not unnatural that the first individual operator to not be correctly captured by the EFM would be one that is less important in saturating the crossing equations, in our case such an operator $\O$ would be characterized (in the $\mathbb Z_2$-even case) by having numerically small OPE coefficients $\lambda_{\phi\phi\O}$ and
 $\lambda_{\phi^2\phi^2\O}$. To assess the situation for the case at hand, we have gathered all known results for the OPE coefficients involving the first few $\mathbb Z_2$-even scalar operators, presented in table~\ref{tab:EvenScalarsOPE}. 
\begin{table}
\caption{OPE coefficients for $\mathbb Z_2$-even scalars. The value of \cite{Padayasi:2021sik} is a conjecture, see \eqref{eq:OPEconjecture}.
}\label{tab:EvenScalarsOPE}
\centering
\begin{tabular}{|c|c|lr|lr|}
\hline
 $i$ & $\O$ & $\lambda^2_{\phi\phi\O}$ & & $\lambda^2_{\phi^2\phi^2\O}$ &
\\\hline
 1 & $\phi^2$  & $2-\frac23\eps$& $(\eps^4)$\cite{Carmi:2020ekr} & $8-8\eps$ & $(\eps^1)$\cite{Henriksson:2020jwk}
\\
 2 &  $\phi^4$ & $\frac{\varepsilon^2}{54} -\frac{47\eps^3}{1458}$ &  $(\eps^3)$\cite{Carmi:2020ekr}   & $6-\frac{20\eps}3$ &  $(\eps^1)$\cite{Bertucci:2022ptt}
\\
 3 & $\phi^6$ & $\frac{5\varepsilon^4}{104796} $ &  $(\eps^4)$\cite{Codello:2017qek}&  $O(\eps^2)$ &
\\
 4 &  $\square^2\phi^4$ & $O(\eps^4)$ & \cite{Alday:2017zzv} & $\frac{1}{6}-\frac{259 \epsilon }{1296}$  & $(\eps^1)$\cite{Bertucci:2022ptt}
\\
 5 & $\phi^8$ & $\frac{7 \eps^6}{37791360}$& $(\eps^6)$\cite{Padayasi:2021sik}* & $O(\eps^4) $ &
\\
 6 &  $\square^3\phi^4$ & $O(\eps^4)$ & & $\frac{1}{175}-\frac{71 \eps}{21000}$ &  $(\eps^1)$\cite{Bertucci:2022ptt}
\\\hline
\end{tabular}
\end{table}

Studying the values in table~\ref{tab:EvenScalarsOPE} we note that all operators of ``$\phi^4$-type'' have $O(1)$ OPE coefficients $\lambda^2_{\phi^2\phi^2\O}$. Therefore these operators can be seen as important in saturating crossing also at small but finite $\eps$. Operators with more than four fields are suppressed in the $\phi^2\times\phi^2$ OPE by powers of $\eps$, unfortunately there are no numerical estimates for the coefficients in these expansions. 

In the $\phi\times\phi$ OPE, a bit more is known, and here all OPE coefficients except that of $\phi^2$ are suppressed by powers of $\eps$. One can think of these as extra powers of $\eps$ as powers of the coupling constant needed to create more fields in the operators appearing in the OPE.
 Interestingly, we also note that operators of increasing powers of $\phi$ have OPE coefficients $\lambda_{\phi\phi\O}^2$ that are not only suppressed by $\eps$ powers, but also contain numerically small constants at the leading non-zero term. In \cite{Padayasi:2021sik}, a formula was conjectured for the OPE coefficient of $\phi^{2k}$ in the $\phi\times\phi$ OPE,\footnote{The values at $k=2$ and $k=3$ have been independently computed by other methods, see references in table~\ref{tab:EvenScalarsOPE}.}
\begin{align}
\label{eq:OPEconjecture}
\lambda^2_{\phi\phi\phi^{2k}}&=\frac{10368k(2k-1)^2\Gamma(k)^4}{6^{4k}\Gamma(2k)}\eps^{2k-2}\,,
\end{align}
which evaluates to $
\frac{\eps ^2}{54},\frac{5 \eps^4}{104976},\frac{7 \eps^6}{37791360},\frac{\eps^8}{952342272}$ for $  k=2,3,4,5$, etc.
We think that it would not be unreasonable for this suppression to survive for finite values of $\varepsilon$.

Let us now compare with our results at finite $\eps$ in figure~\ref{fig:even0}. Among the first six operators, the $\phi^8$ operator that is not seen in the numerics is the one with the largest number of fields, which coincides with being the operator with smallest OPE coefficients reported in table~\ref{tab:EvenScalarsOPE}. Where explicit formulas are available, the suppression holds both perturbatively in $\eps$, and numerically in the size of the constants.

\subsection{Other operator spectra}

\begin{figure}
\centering
\includegraphics[width=0.95\textwidth]{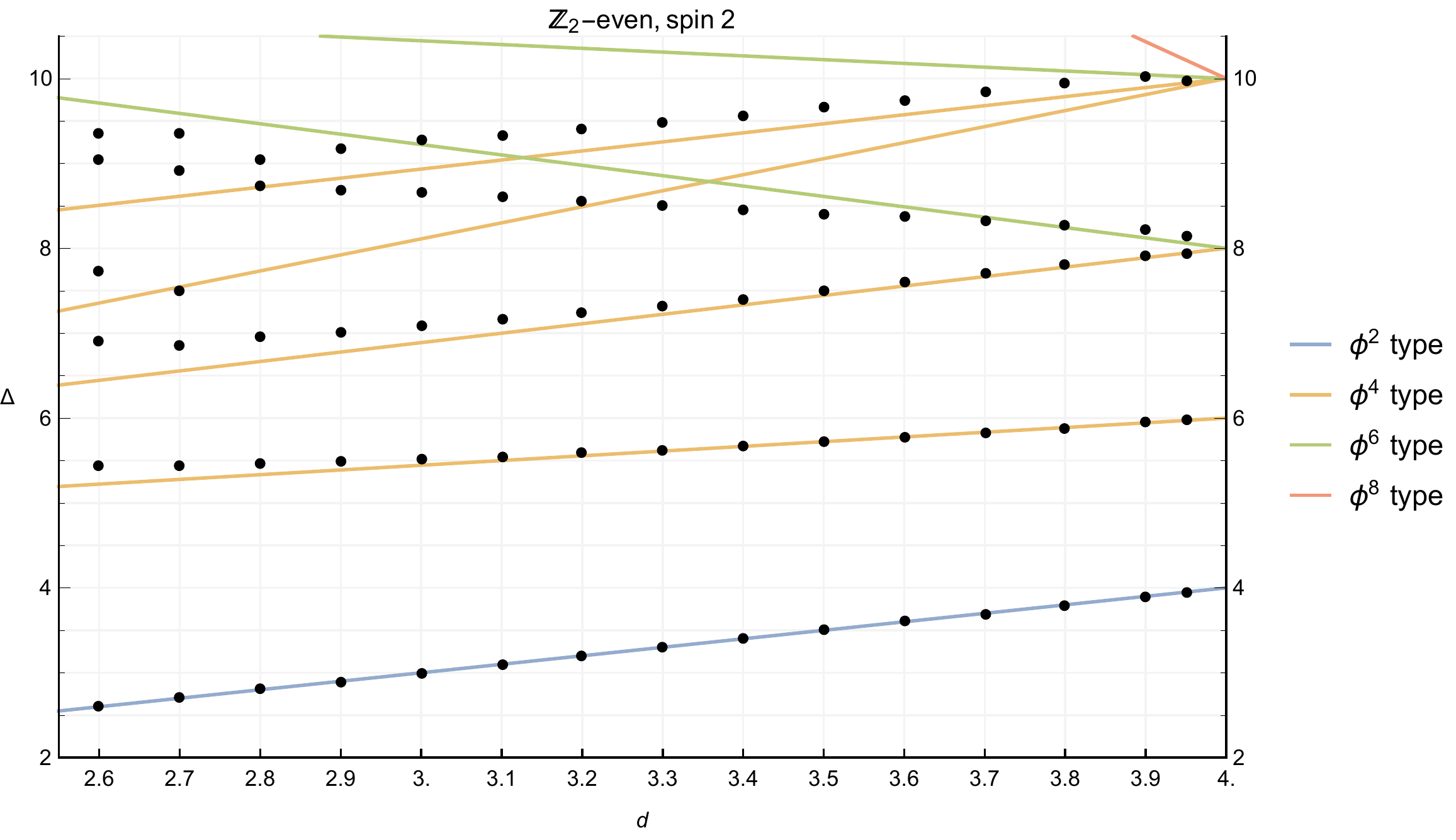}
\caption{$\mathbb Z_2$-even operators of spin 2.}\label{fig:even2}
\end{figure}

\begin{figure}
\centering
\includegraphics[width=0.95\textwidth]{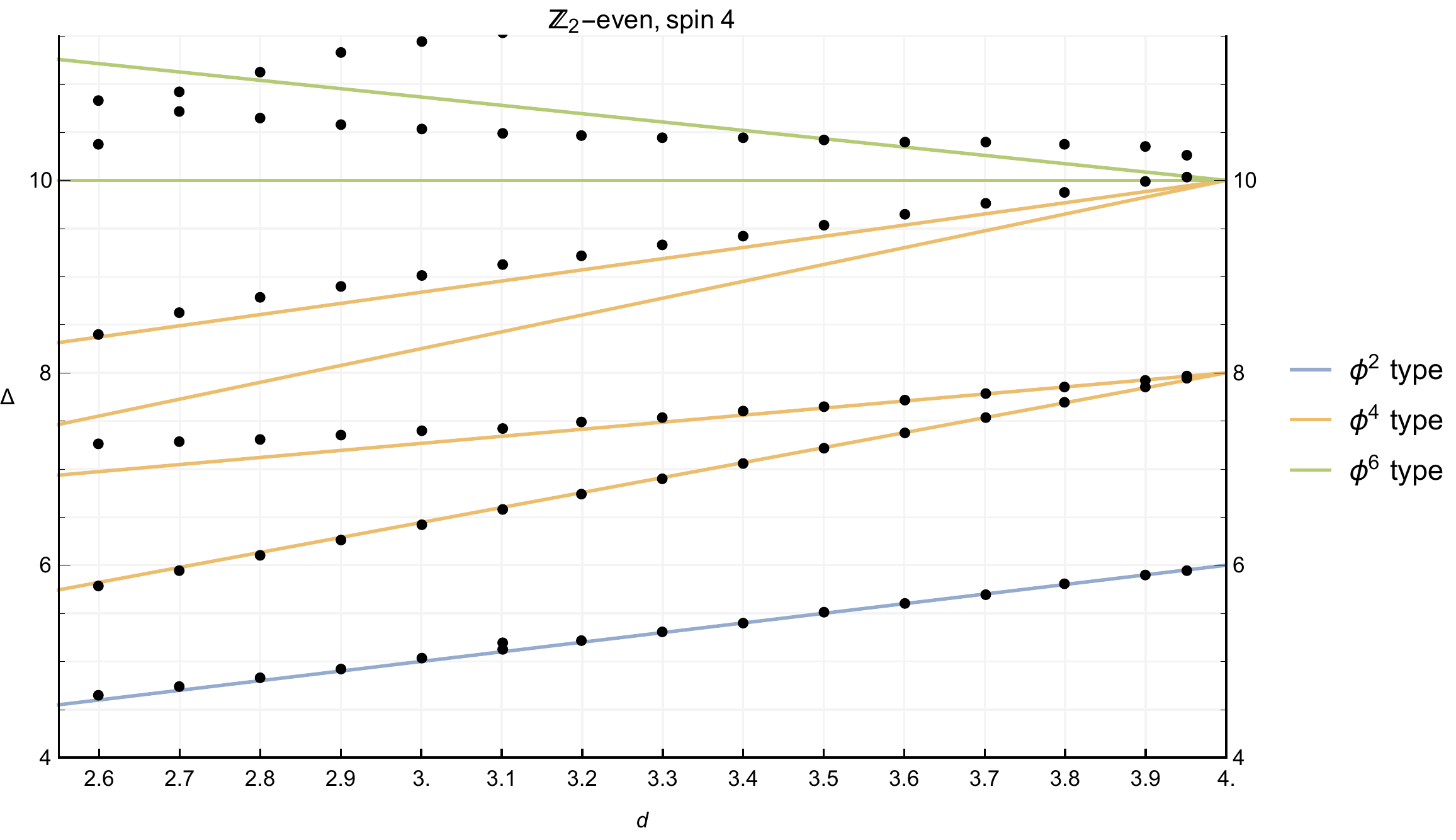}
\caption{$\mathbb Z_2$-even operators of spin 4.}\label{fig:even4}
\end{figure}

\begin{figure}
\centering
\includegraphics[width=0.8\textwidth]{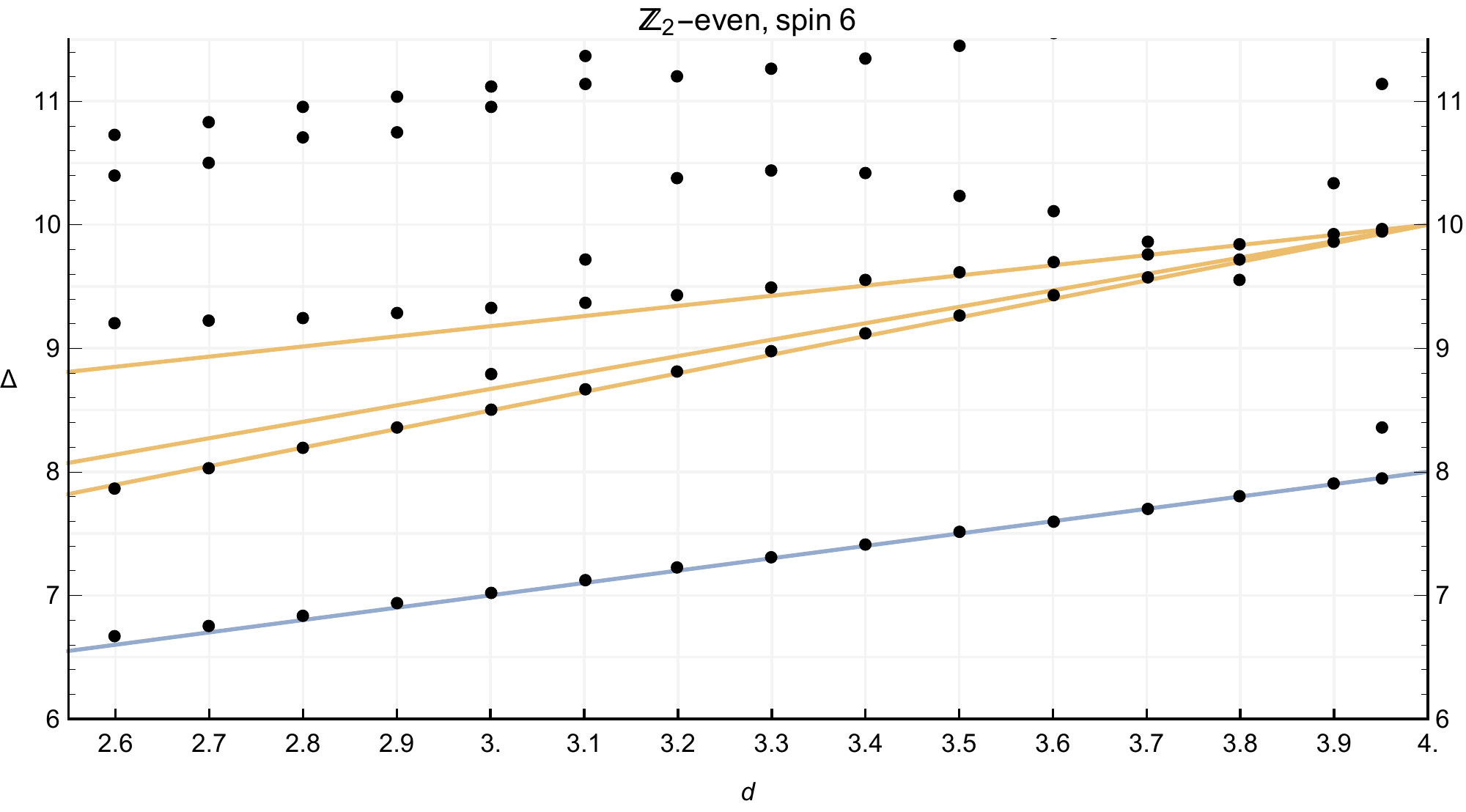}
\caption{$\mathbb Z_2$-even operators of spin 6.}\label{fig:even6}
\end{figure}

\begin{figure}
\centering
\includegraphics[width=0.8\textwidth]{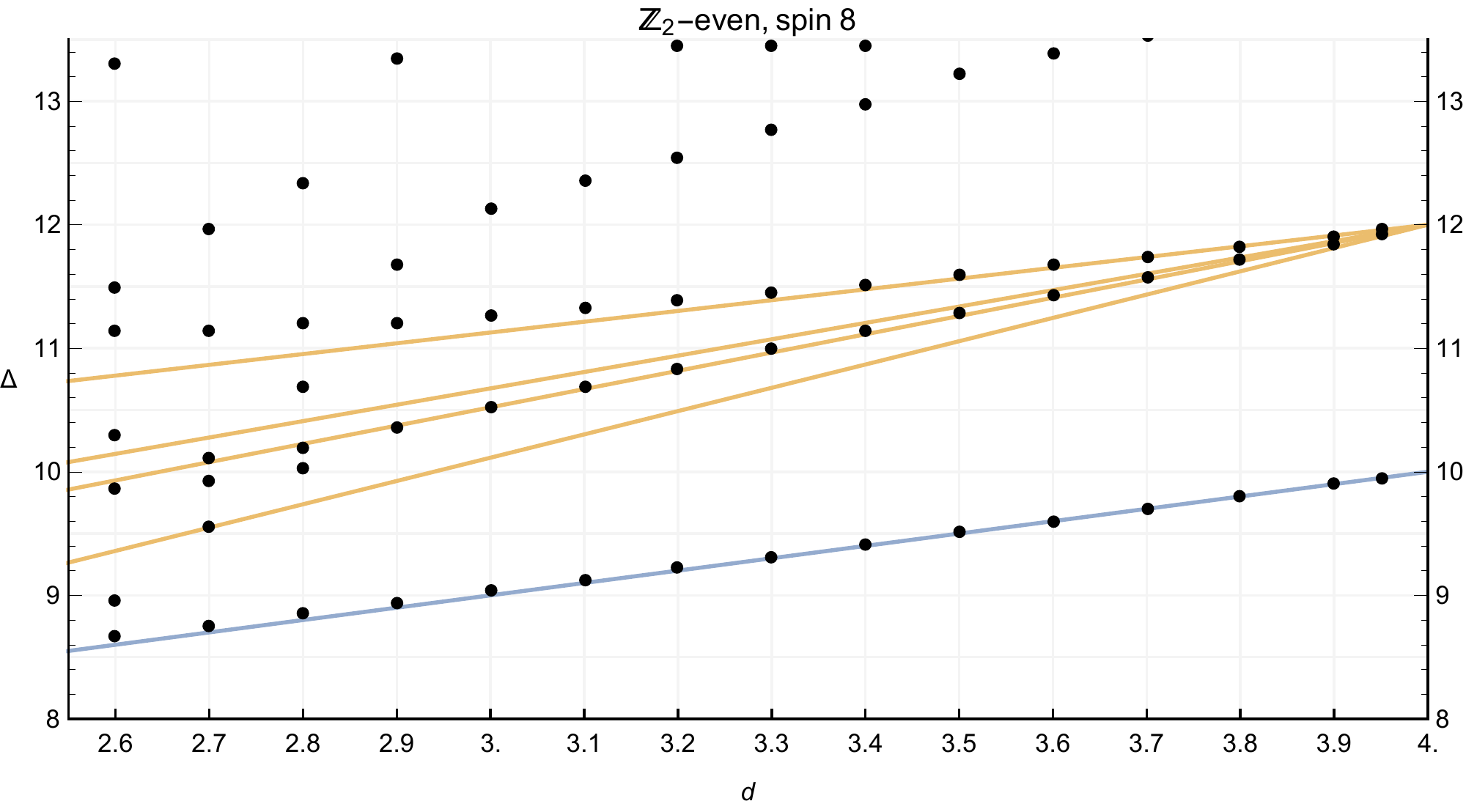}
\caption{$\mathbb Z_2$-even operators of spin 8.}\label{fig:even8}
\end{figure}

\begin{figure}
\centering
\includegraphics[width=0.8\textwidth]{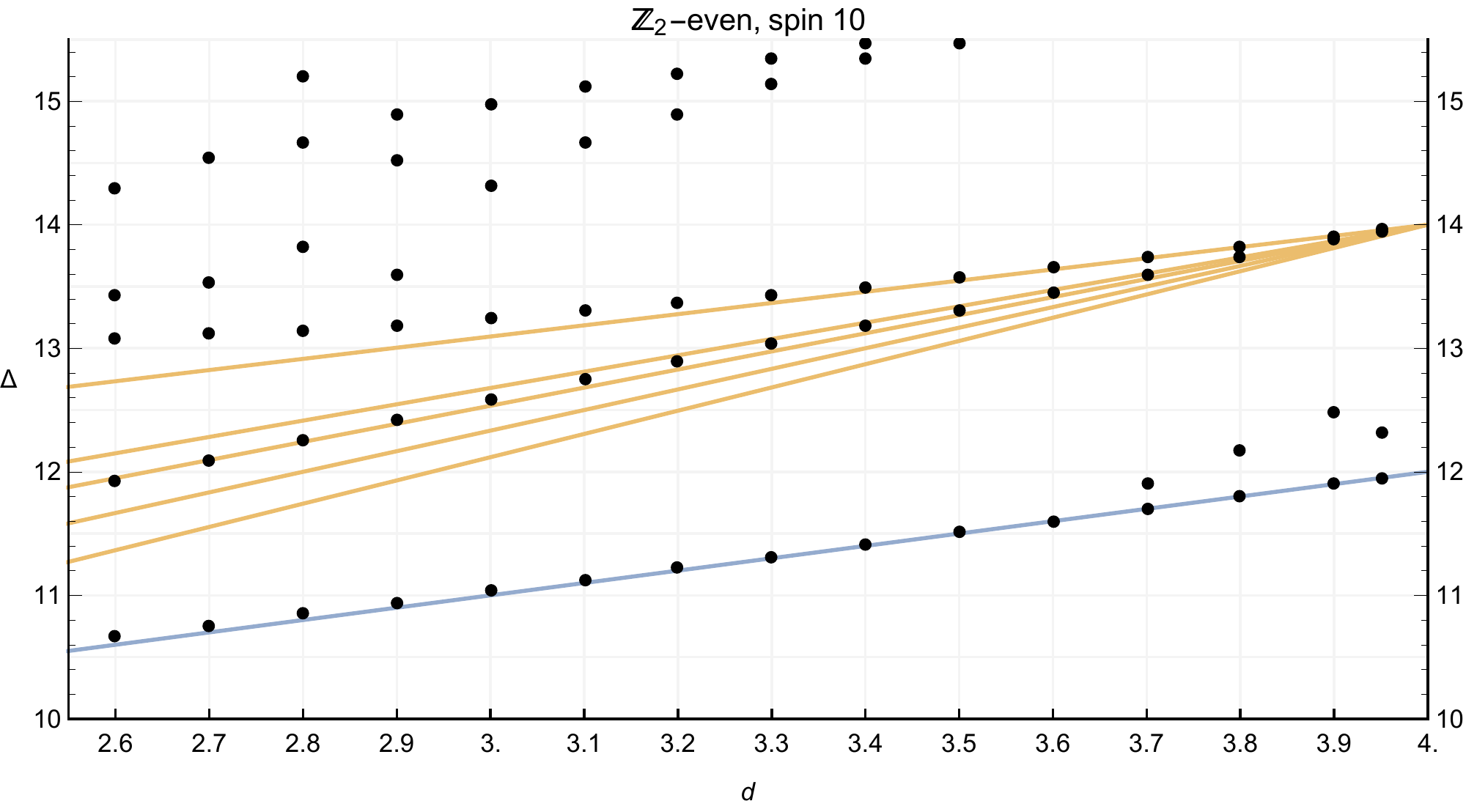}
\caption{$\mathbb Z_2$-even operators of spin 10.}\label{fig:even10}
\end{figure}

\begin{figure}
\centering
\includegraphics[width=0.8\textwidth]{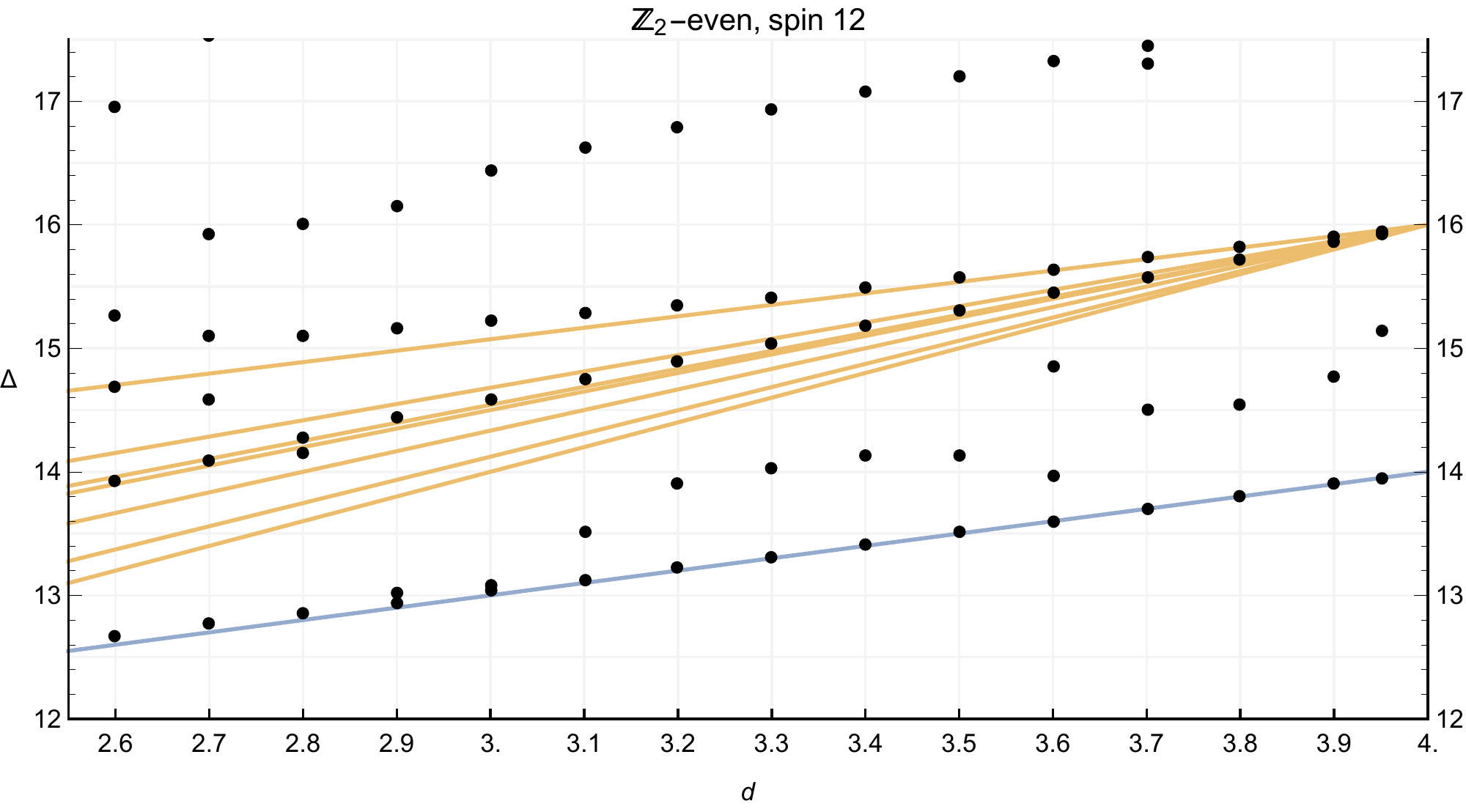}
\caption{$\mathbb Z_2$-even operators of spin 12.}\label{fig:even12}
\end{figure}

\begin{figure}
\centering
\includegraphics[width=0.95\textwidth]{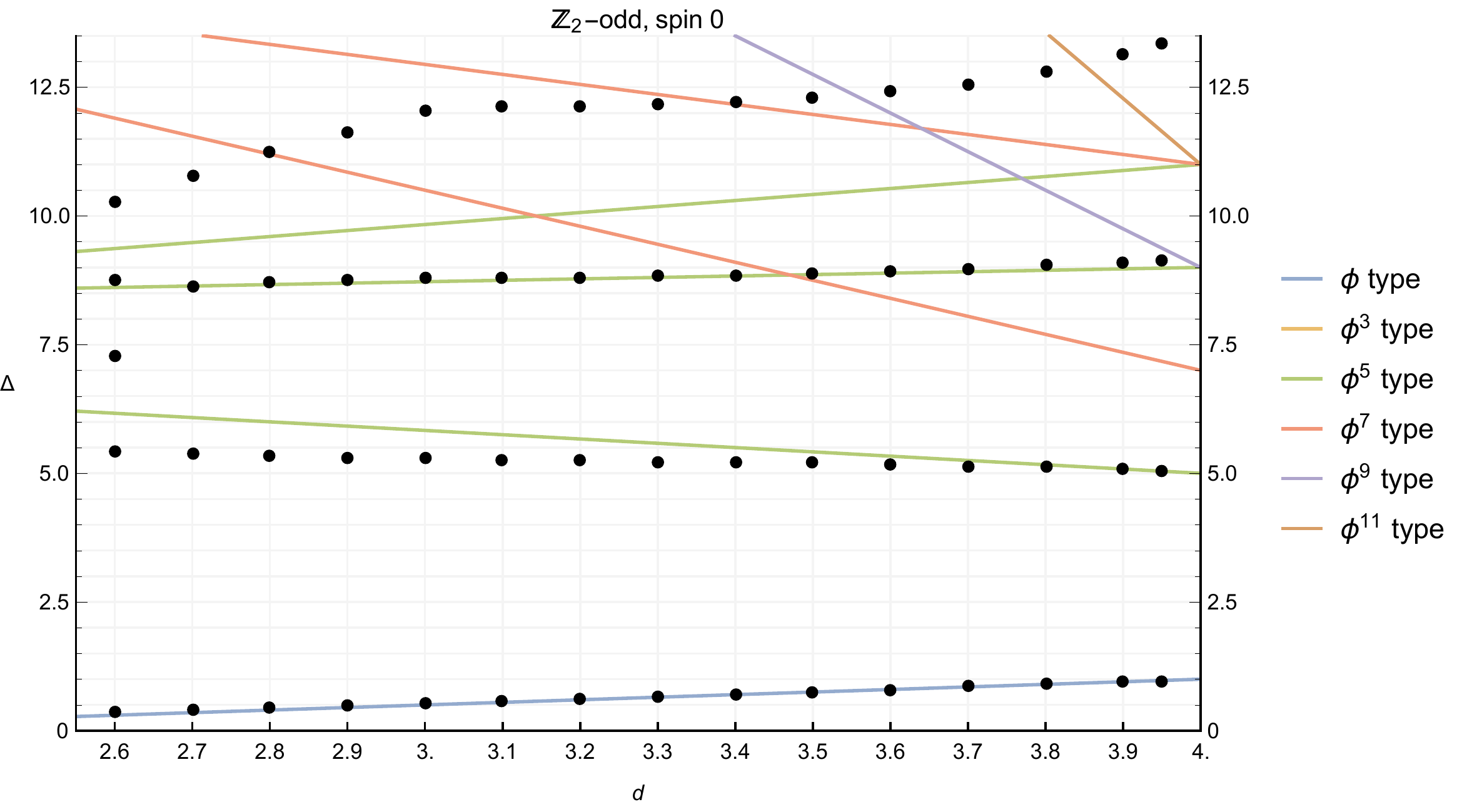}
\caption{$\mathbb Z_2$-odd operators of spin 0.}\label{fig:odd0}
\end{figure}

\begin{figure}
\centering
\includegraphics[width=0.95\textwidth]{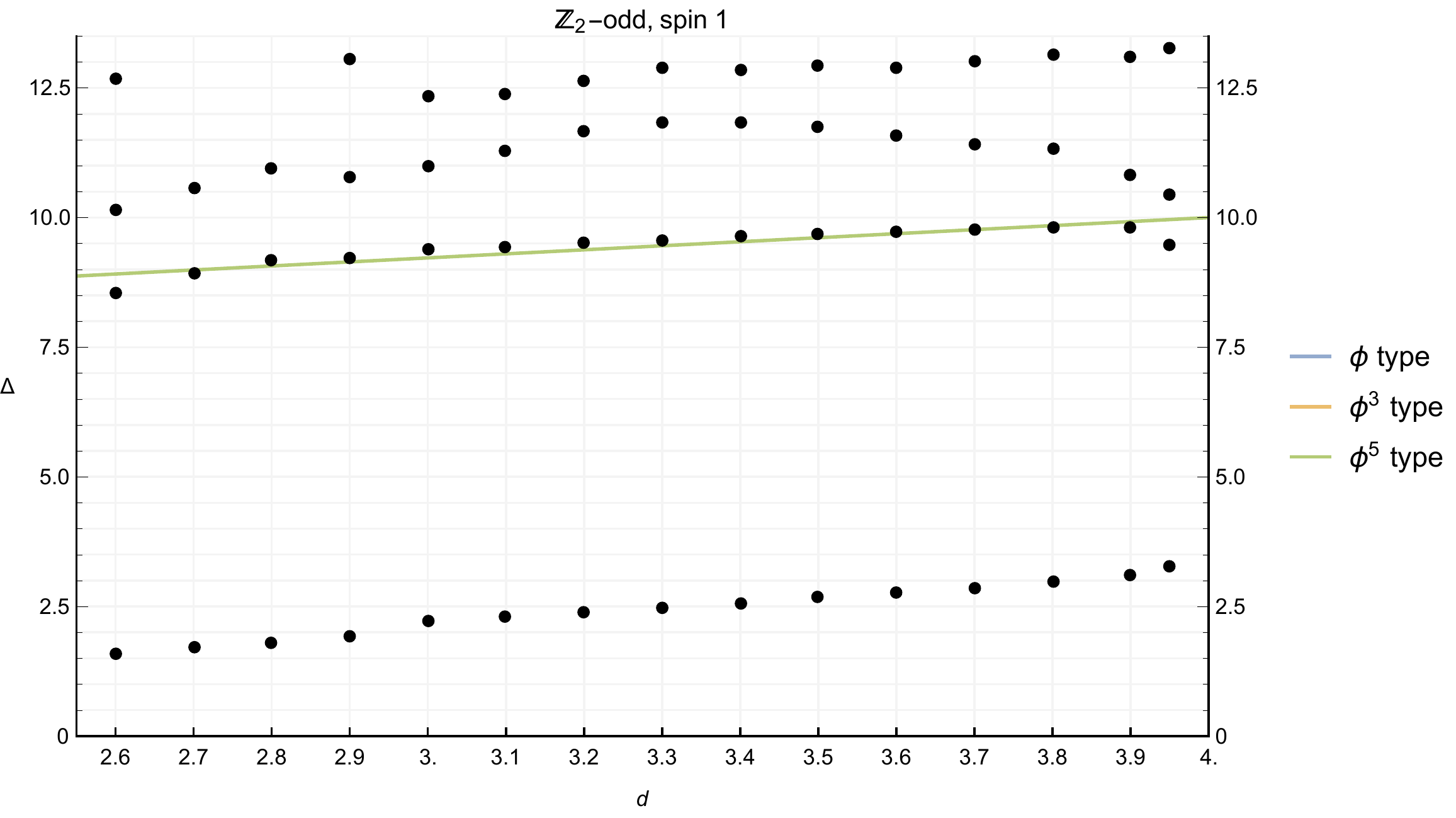}
\caption{$\mathbb Z_2$-odd operators of spin 1.}\label{fig:odd1}
\end{figure}

\begin{figure}
\centering
\includegraphics[width=0.95\textwidth]{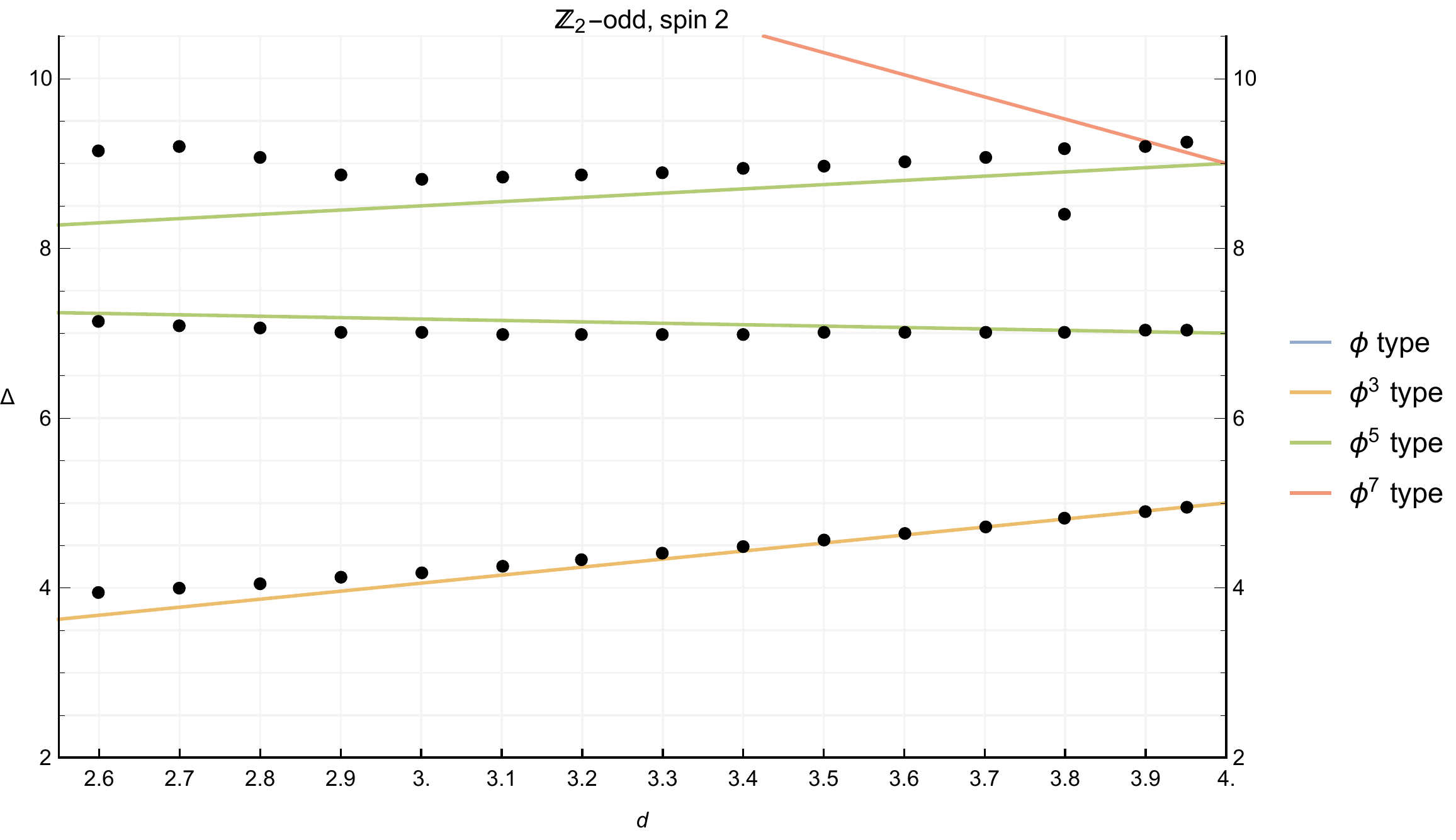}
\caption{$\mathbb Z_2$-odd operators of spin 2.}\label{fig:odd2}
\end{figure}

\begin{figure}
\centering
\includegraphics[width=0.95\textwidth]{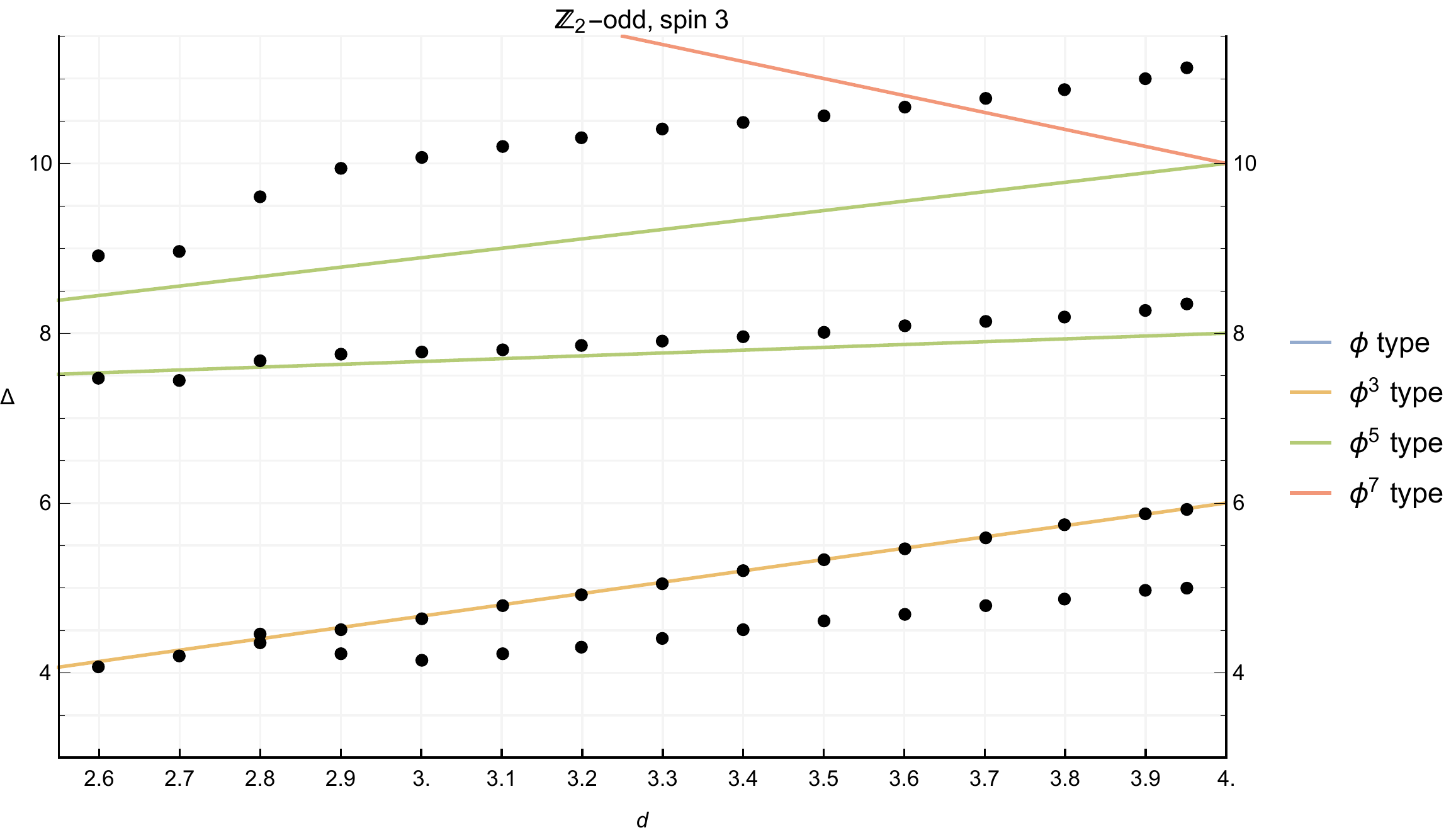}
\caption{$\mathbb Z_2$-odd operators of spin 3.}\label{fig:odd3}
\end{figure}

\begin{figure}
\centering
\includegraphics[width=0.8\textwidth]{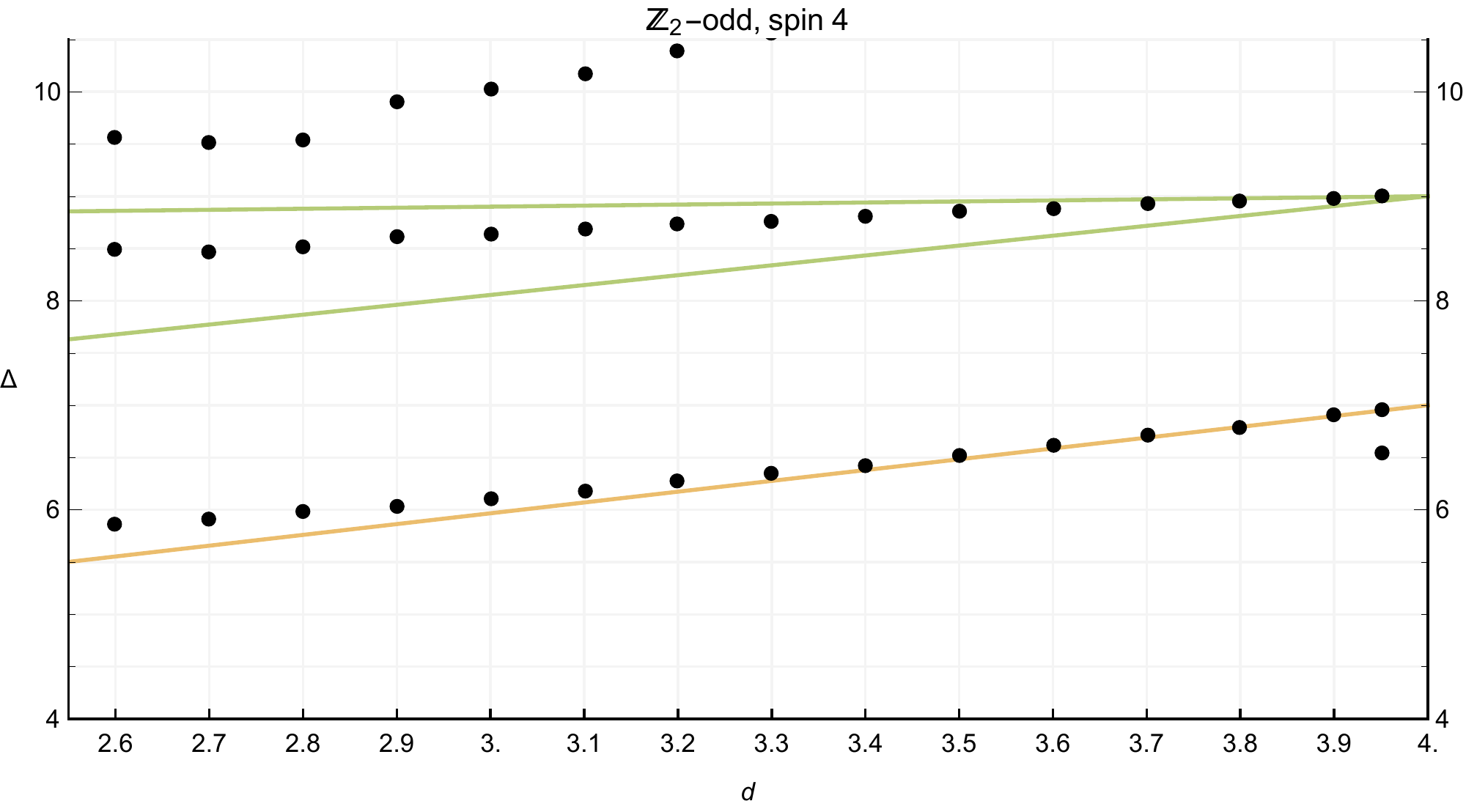}
\caption{$\mathbb Z_2$-odd operators of spin 4.}\label{fig:odd4}
\end{figure}

\begin{figure}
\centering
\includegraphics[width=0.8\textwidth]{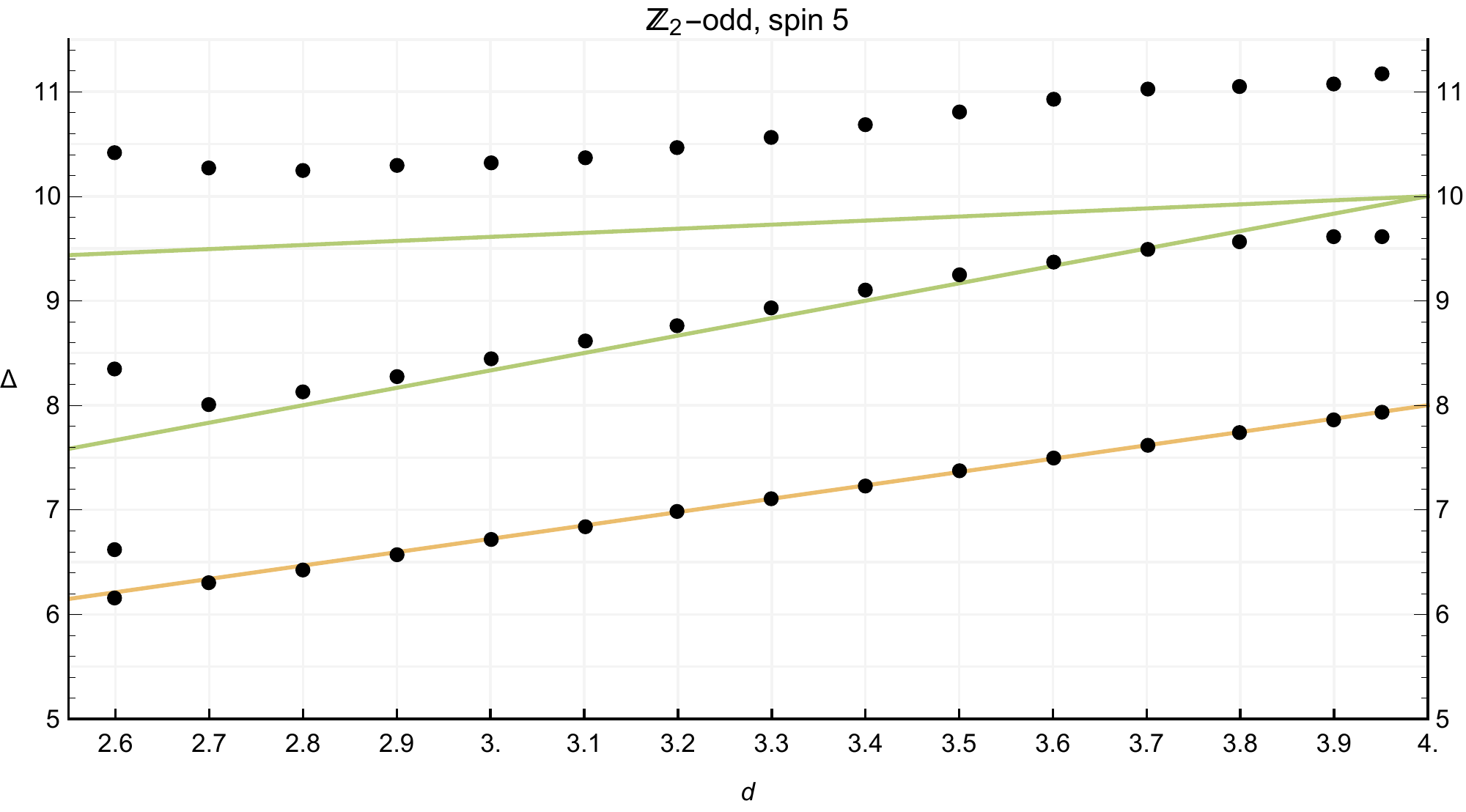}
\caption{$\mathbb Z_2$-odd operators of spin 5.}\label{fig:odd5}
\end{figure}

\begin{figure}
\centering
\includegraphics[width=0.8\textwidth]{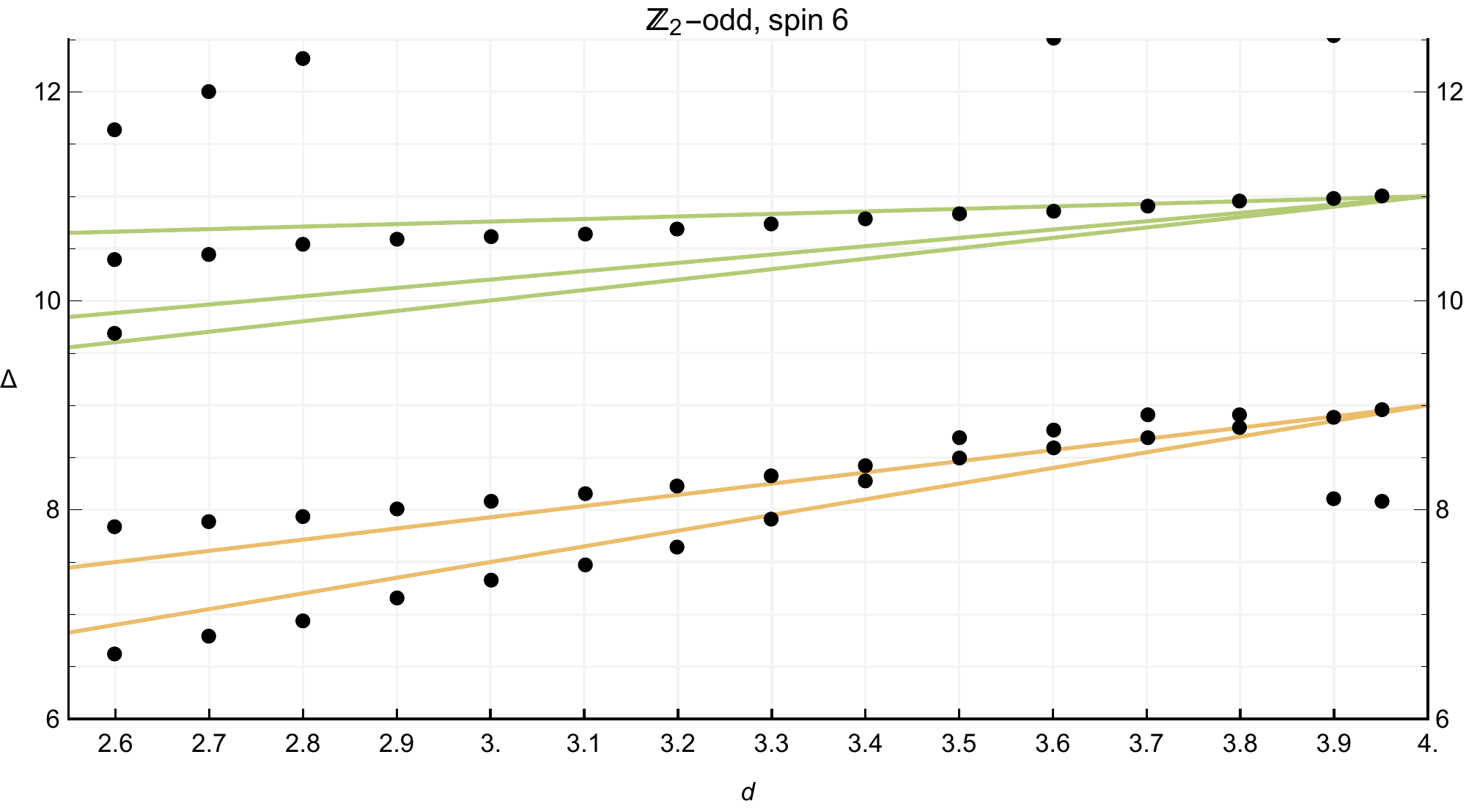}
\caption{$\mathbb Z_2$-odd operators of spin 6.}\label{fig:odd6}
\end{figure}

Having discussed the $\Z_2$-even scalars in detail, we now move on to consider different symmetry irreps. 
In figures \ref{fig:even2}--\ref{fig:even12} we give the operator spectra for $\Z_2$-even operators from spin-$2$ up to spin-$12$ with a step of $2$. Since the $\Z_2$-even operators in our crossing system only appear as the exchange of identical scalars, we cannot access any $\Z_2$-even operator of odd spin in this work.\footnote{However, these operators would be exchanged in a system of correlators that includes two different $\Z_2$-even external operators, for instance by adding $\phi^4$ or the stress-tensor $T^{\mu\nu}$ to the system.}
Similarly, in figures \ref{fig:odd0}--\ref{fig:odd6} we exhibit the spectra for $\Z_2$-odd operators from spin $0$ to spin $6$ with a step of $1$. In each of the plots we display the result of our computation together with order $\eps$ anomalous dimensions for operators in the $\eps$-expansion. We display a subset of this perturbative data is collected in tables~\ref{tab:EvenEps} and \ref{tab:OddEps}.

In general, we find rather good agreement between some of the low-lying operators in perturbation theory, and some sequences of points in the numerical data. Whenever there is a single sequence of points that approach a single line starting at $d=4-\eps$, we can conclude that this set of non-perturbative data would be an estimate of the scaling dimension of that particular operator at finite $\eps$. Also in a couple cases with two-fold degeneracy at $\eps\to0$, the identification of operators between numerics and perturbation theory seems unambiguous, namely the as the dimension $8+O(\eps)$ $\Z_2$-even operators at spins 2 and 4 (figures~\ref{fig:even2} and \ref{fig:even4}).

 In a couple of our plots, odd spin 1 and spin 3 (figures~\ref{fig:odd1} and \ref{fig:odd3}), there are persistent sequences of minima of the extremal functional that do not correspond to operators in the $\eps$-expansion. We have checked that they are not sporadic features of a special navigator point but instead are rather persistent features just like the dimensions that correspond to physical operator.\footnote{Figure~\ref{fig:even6} for $\Z_2$-even operators of spin 6 shows an interesting feature. Around $d\sim 3.6$, it looks like one of the three operators with dimension $\Delta=10+O(\eps)$ appears to deviate from the perturbative line and migrate upwards as $d$ decreases. Then around $d=3.1$ it looks like the same operator moves down again to appear below the top yellow line at $d=3$. This behavour is interesting, since \cite{Simmons-Duffin:2016wlq} reported only two operators in this region. However, a closer analysis of several spectra before the final Navigator point shows that the points $(3,8.78)$ and $(3.1,9.72)$ are unstable.}
However, we do not believe that they are signals of new operators that were for some reason not previously discovered in perturbation theory.\footnote{The perturbative spectrum reported in \cite{Kehrein:1994ff} has been independently investigated recently by one of us \cite{Henriksson:2022rnm}, and we have no reason to believe it to be incorrect. Moreover, for the case of figure~\ref{fig:odd3}, it is not possible to construct a $\Z_2$-odd spin-3 operator with dimension $5+O(\eps)$ out of fields and derivatives in the $\eps$-expansion.}

In summary, our results seem to confirm that the $\varepsilon$-expansion is quite accurate at least in the Ising case. Also, at least for the operators that we can track relatively well, there does not seem to be any operator that decouples from the spectrum for some value of $d$. To be more precise, by being able to track an operator well, we refer to the scenario were both perturbation theory and the bootstrap overlap well at small $\varepsilon$, and then the corresponding curve evolves smoothly. As a counterexample, in figure~\ref{fig:even10} close to $d=4$ we see a few cases of an extra operator close to $\Delta=12$, however there is no corresponding prediction for any extra operator in the $\varepsilon$-expansion, we thus conclude that these points are spurious.

\subsection{OPE coefficients}

Let us conclude this section by displaying the values of some OPE coefficients in the theory. The data for OPE coefficients follow from the extremal crossing solution, which we extract by \texttt{spectrum.py} \cite{spectrum}. This method becomes quite unstable for OPE coefficients higher up in the spectrum, even when there are clear minima of the extremal functional. We therefore focus on the leading two OPE coefficients. In the numerical problem, they correspond to the OPE coefficients involving the leading two operators $\sigma$ and $\epsilon$, which by the above considerations successfully matched with the operators $\phi$ and $\phi^2$ in the $\eps$-expansion. Therefore we have

\begin{figure}
\centering
\includegraphics[width=0.94\textwidth]{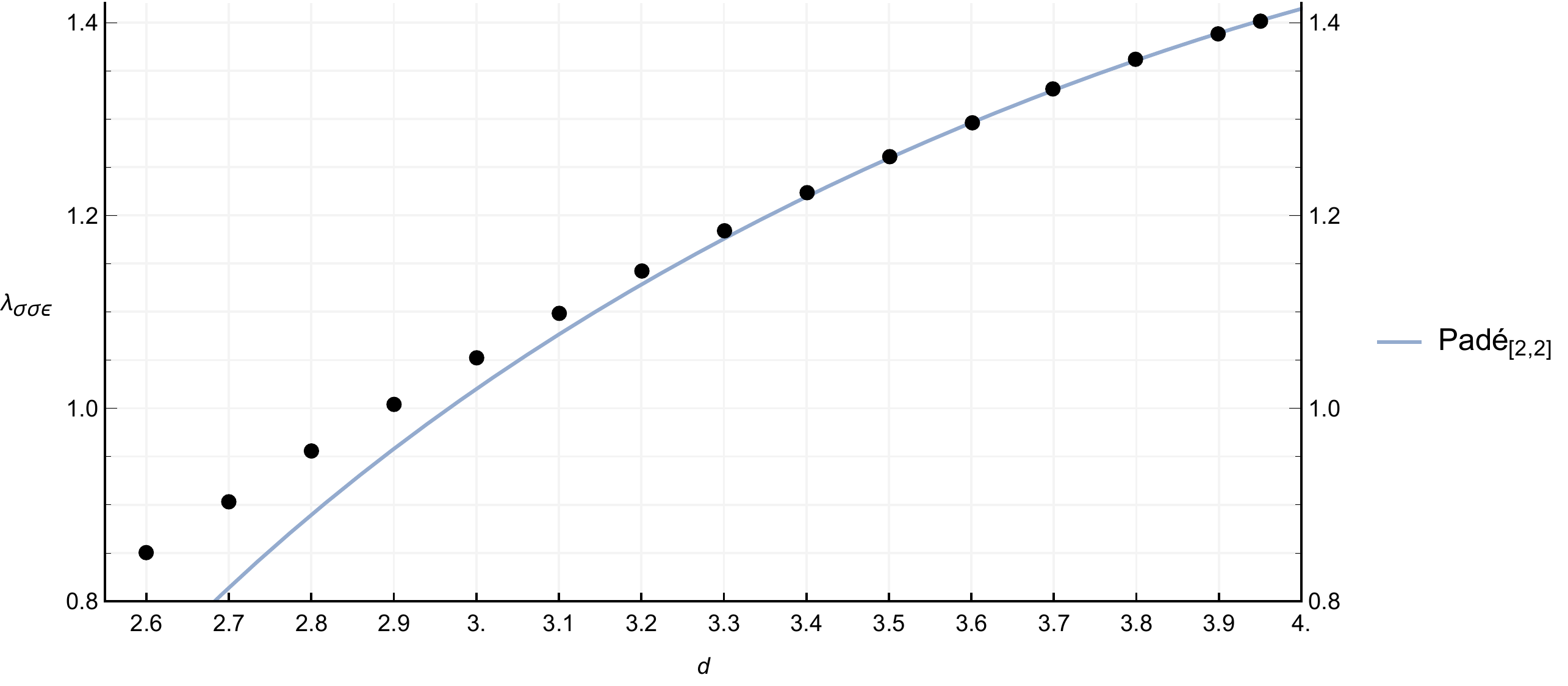}
\caption{OPE coefficient $\lambda_{\phi\phi\phi^2}$.}\label{fig:opeffs}
\end{figure}
\begin{figure}
\centering
\includegraphics[width=\textwidth]{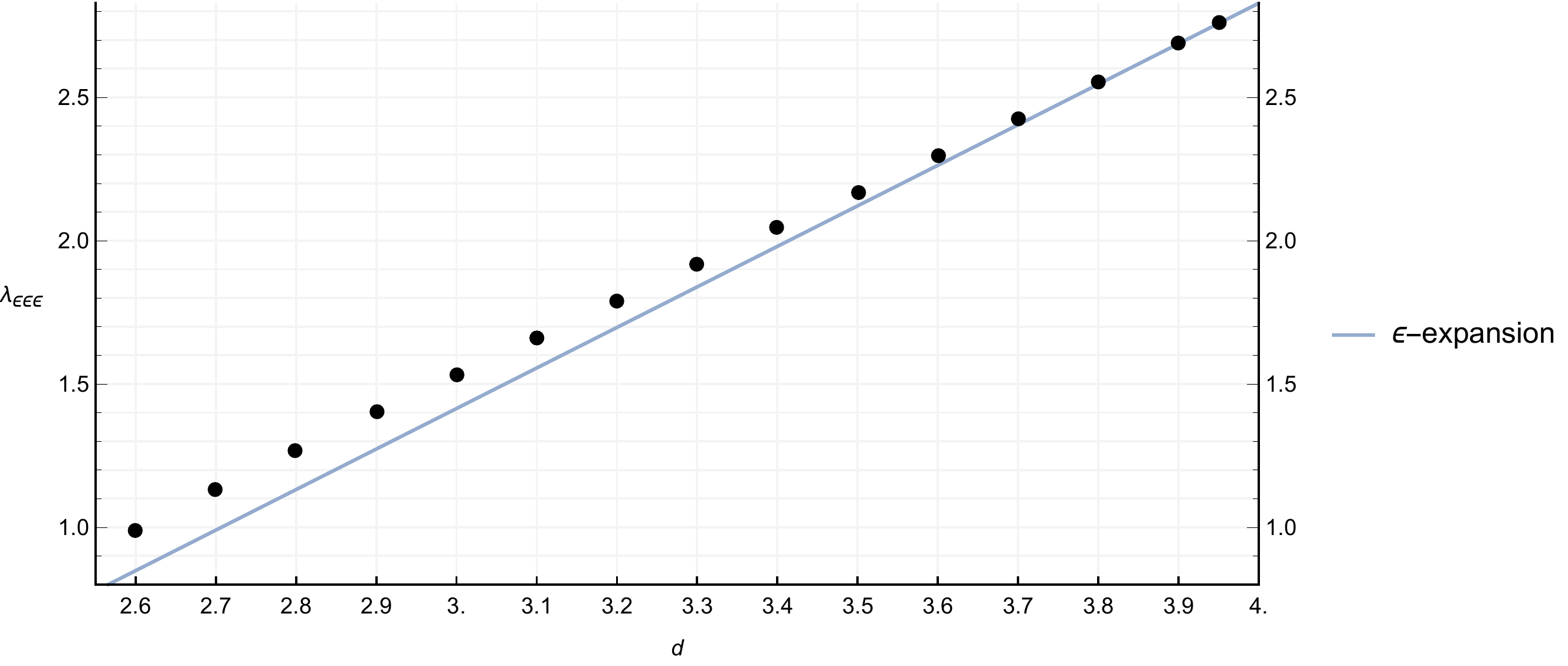}
\caption{OPE coefficient $\lambda_{\phi^2\phi^2\phi^2}$.}\label{fig:opesss}
\end{figure}
\begin{equation}
\lambda_{\sigma\sigma\epsilon}\leftrightarrow\lambda_{\phi\phi\phi^2}, \qquad 
\lambda_{\epsilon\epsilon\epsilon}\leftrightarrow\lambda_{\phi^2\phi^2\phi^2}.
\end{equation}
In figures~\ref{fig:opeffs} and \ref{fig:opesss} we display our results for these OPE coefficients extracted at the navigator minimum.\footnote{We can only extract these up to an overall sign since they appear squared in the crossing equations. Matching with the $\eps$-expansion fixes the sign to be positive.} The plots show good agreement with the perturbative values. For the first OPE coefficients, we compare with a Pad\'e approximant constructed using the order $\eps^4$ result
\begin{equation}
\lambda_{\phi\phi\phi^2}=\sqrt2-0.2357 \eps-0.1680 \eps^2+0.1036 \eps^3-0.2248 \eps^4+O(\eps^5)
\end{equation}
from \cite{Carmi:2020ekr}.\footnote{The terms up to order $\eps^3$ are known on exact form and were computed in \cite{Gopakumar:2016cpb}. The order $\eps^4$ term is only known numerically.} For the second OPE coefficient, only the linear correction is known \cite{Henriksson:2020jwk},
\begin{equation}
\lambda_{\phi^2\phi^2\phi^2}=\sqrt8-\sqrt2\eps+O(\eps^2).
\end{equation}
Finally, we can compare our estimates in $d=3$ with the rigorous values in \cite{Kos:2016ysd},
\begin{align}
\label{opecomparisons}
(\lambda_{\sigma\sigma\epsilon},\lambda_{\epsilon\epsilon\epsilon})\big|_{\mathrm{min}}&=(1.0518527,\quad\ \ \, 1.5324484),
\\
  (\lambda_{\sigma\sigma\epsilon},\lambda_{\epsilon\epsilon\epsilon})\big|_{\text{rigorous}}&=(1.0518537(41),1.532435(19)).
\label{opecomparisons2}
\end{align}
We see that our values are inside the rigorous error bars.

\section{Discussion and outlook}
In this work we performed a comprehensive conformal bootstrap study on the low-lying spectrum of the Ising CFT. This was done for a mixed correlator system involving all possible correlators with $\sigma$ and $\epsilon$ as externals. With this system we were able to access $\Z_2$-even operators of even spin, and $\Z_2$-odd operators of any spin. We explicitly presented results for $\Z_2$-even operators up to spin $12$ and $\Z_2$-odd operators up to spin $6$ obtained using the extremal functional method. 
The main observation of our work is that, in absence of mixing effects, the $\varepsilon$-expansion produces results that agree well with our non-perturbative results for various subleading operators. 
This good agreement is somewhat surprising, since in many cases we only had access to one order in the perturbative expansion.
In the case of mixing, the above statement holds away from the mixing region. In the mixing region, on the other hand, we find that the operator dimensions match well with a fitted hyperbola \eqref{eq:fittedHyperbolaEigs}, which is expected to give a good local approximation of the level repulsion caused by operator mixing.
An interesting goal for the future would be to reproduce the curve in figure~\ref{fig:repulsion} with an analytic computation, in particular to give an estimate for the (half) minimal distance $x=0.067957$.

Another observation is that in the $\Z_2$-even spin-$0$ channel, we were able to track the first four operators seemingly quite reliably. I.e. we found good agreement as $d\rightarrow 4$, and then smooth continuous lines with no apparent spurious operators in between (at least down to $d=3$). We note similarly good agreement for some leading operators in the other representations, and we can therefore see our work as an extension of \cite{Cappelli:2018vir} to include operators with higher twist.
This appears rather promising for future studies, where with higher precision we may be able to study operators even higher up in the spectrum. 

There are a few obvious ways in which our results can be improved in the future. If we include the operator $\phi^4$ $\sim$ $\epsilon^\prime$ as an external operator in the system of correlators, we will now be able to access $\Z_2$-even operators of odd spin. 
Such operators are inaccessible in the $\sigma$, $\epsilon$ system, since they can only appear in the OPE of non-identical external operators. We can also include external spinning operators, such as the stress tensor, something that will allow us to access space(time) parity odd operators. The aforementioned generalizations would then give access the entire spectrum of the 3d Ising CFT, at least in principle -- in practice we expect to be able to study only a few low lying operators in each sector.
We find this particularly appealing, since it would be one of the few, if any, non-integrable field theories where such an amount of data would be known non-perturbatively.

In this work, as an initial proof of concept we showed evidence for level repulsion phenomenon using the EFM to provide highly accurate estimates. It would be interesting to turn these estimates into rigorous bounds and to determine the relevant curves more precisely. This can be done by including the values for $\Delta_{\epsilon''}$ and $\Delta_{\epsilon'''}$ explicitly in the navigator function \cite{Reehorst:2021hmp}. Then one can minimize the distance between the two curves as an efficient optimization problem.\footnote{We acknowledge conversations with Ning Su on this point.} Enlarging the search space in such a way comes at the cost of increased numerical run-times. However, we expect that with the rapidly evolving bootstrap technology such numerics should be feasible in the not so distant future. Such a study would be expected to conclusively settle between the repulsion scenario \eqref{eq:mixinggen} and the annihilation scenario \eqref{eq:mixinggenAntisymm}.

\begin{figure}
\centering
\includegraphics[width=0.86\textwidth]{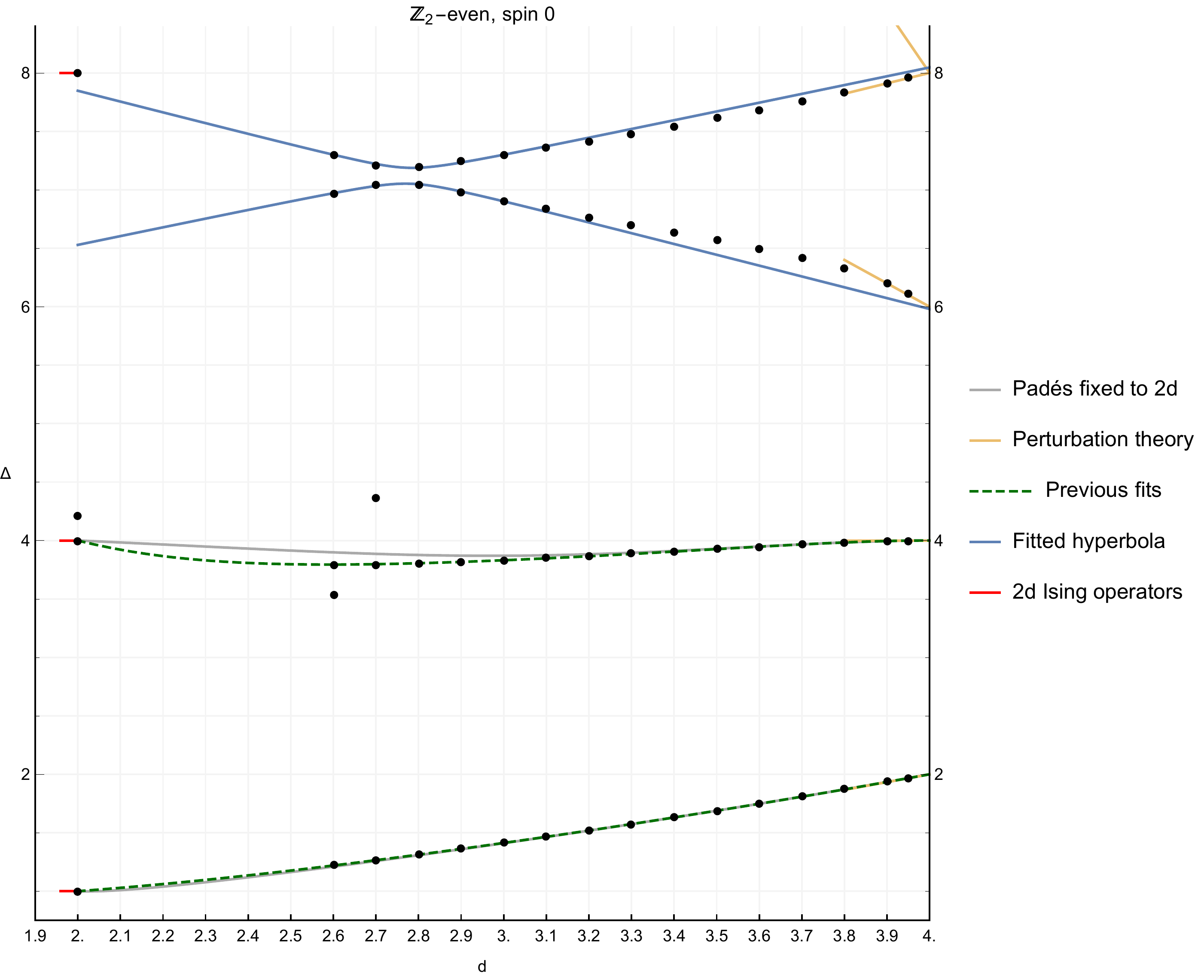}
\caption{Plot showing the collected state-of-the-art picture for the first four $\Z_2$-even scalars, including our preliminary results in $d=2.001$. The dots refer to results from the numerical bootstrap study, the gray lines are Pad\'e approximants computed from the $\eps$-expansion and values in 2d, the green dashed lines to previous fits from \cite{Cappelli:2018vir}, and blue lines to the hyperbola fit from the mixing region, \eqref{eq:fittedHyperbolaEigs}. Red markers denote the position of quasiprimaries in $d=2$. Note that \cite{Cappelli:2018vir} has values for the first two operators for a more dense sampling of values of $d$ near $d=2$.}\label{fig:d2001}
\end{figure}

The Ising CFT is of experimental relevance in spacetime dimensions $d=2$ and $d=3$. The $d=2$ theory has been solved exactly, while the $d=3$ theory remains an active area of research, also from a theoretical point of view. This study focused on the region where the first strong example of level repulsion takes place, but there are also questions to be answered about the $d \rightarrow 2$ behaviour of the CFT.
For example, the $d=2$ theory is to some approximation accessible via the $d=4-\varepsilon$ expansion \cite{LeGuillou:1987ph,Kompaniets:2017yct} and as emphasized before, the connection has been corroborated by bootstrap studies in intermediate dimensions \cite{El-Showk:2013nia,Cappelli:2018vir}.\footnote{ Another interesting observation is that doing perturbation directly in $d=2$, if one goes to high enough order and performs re-summations, one obtains values for critical exponents surprisingly close to the exact result \cite{Serone:2018gjo}.}   See also \cite{Ballhausen:2003gx} for a study in intermediate dimensions using the non-perturbative renormalization group.
Hence, one could try to understand how a theory (Ising in $d=4-\varepsilon$) that seems to not be exactly solvable, continuously deforms into an integrable theory in $d=2$, see \cite{Cappelli:2018vir,Li:2021uki} for work in this direction. It will be interesting for the future to address this limit for a larger set of operator dimensions. 
To this extent, we present an appetizer plot in figure~\ref{fig:d2001}, where we include some preliminary results from performing our bootstrap study in $d=2.001$. In this figure we show Pad\'e approximants and the interpolating curves from \cite{Cappelli:2018vir} for the first two operators, and our hyperbola for the following two operators. Na\"ively, the figure suggests that the third scalar singlet decouples from the spectrum of the theory as we approach $d=2$. 
We plan to investigate this phenomenon more carefully in future work.

There are other theories with an interpolating parameter, whose spectrum would be interesting to study with respect to spectrum continuity and level repulsion. An immediate target is the generalization to the $O(N)$ CFT, which contains Ising at $N=1$. 
A first step in this direction was taken in \cite{Sirois:2022vth}, where a portion of the spectrum was probed as a function of $N$ and the dimensionality $d$, focusing on the region $d\geqslant 3$. An interesting open question in this context is the ``Cardy--Hamber line'' starting at $N=d=2$, across which scaling dimensions are conjectured to be non-analytic \cite{Cardy:1980at}. A recent study using non-perturbative RG found no evidence of non-analyticity \cite{Chlebicki:2020pvo}, and it would be interesting to investigate this with the bootstrap. With respect to level repulsion, the large $N$ expansion of the $O(N)$ model shows some interesting features of level repulsion that may be interesting to study numerically \cite{Derkachov:1998js}.\footnote{Specifically, the order $1/N$ anomalous dimensions of the third and the fourth singlet scalars coincide at $d=17 - \sqrt{193}=3.1076$. The apparent contradiction with the non-crossing rule is resolved by taking higher-order corrections into account. At order $1/N^2$, the anomalous dimension has only been determined for fourth operator, and it has a pole exactly at the coincident point. It would be interesting to study this effect non-perturbatively.} Also the level non-crossings in $\mathcal N=4$ SYM studied by \cite{Korchemsky:2015cyx} and mentioned in section~\ref{sec:repulsion} would be interesting target for a non-perturbative bootstrap study. Here the level repulsion between Konishi and the first double-trace operator would be observable in a study at large but finite $N_c$, and at finite coupling. With the recent localization techniques for inserting the value of the coupling constant into the bootstrap system \cite{Chester:2021aun}, this problem looks tractable.

Another seemingly straightforward generalization of our work would be to study the supersymmetric Ising CFT. In this case the CFT can be found as the fixed point of the simplest GNY model one may write down \cite{Fei:2016sgs}. 
In that case, the supersymmetry appears to be emergent, even if one does not impose it from the start. At the level of the $\varepsilon$-expansion, the signal for this is the integer spacing of specific scaling dimensions. As was shown in \cite{Rong:2018okz,Atanasov:2018kqw} it is possible to get an island in parameter space using exactly the same correlator system we used for this work. 
Using the same system of crossing equations, the authors were also able obtain high precision estimates for CFT data \cite{Atanasov:2022bpi}. 
One potential complication in this direction is that evanescent operators, i.e. the operators responsible for violations of unitarity in fractional dimensions as discussed in \cite{Hogervorst:2015akt}, may appear for smaller values of the scaling dimensions in fermionic theories than in purely bosonic theories \cite{Hogervorst:2015akt,Dugan:1990df,Gracey:2008mf,Ji:2018yaf}. Hence, if these violations become sizeable enough to be detected, this may have an effect on the bootstrap.

One may also study more complicated multi-scalar theories with various global symmetries. A number of these are interesting experimentally, and others are interesting due to field theoretical considerations.\footnote{See \cite{Mukamel1975,Mukamel1976,Mukamel1976b,Bak1976,Pelissetto:2000ek} among others.} For example, once one allows for a global symmetry that is more complicated, various phenomena can be observed, such as annihilation of fixed points, or fixed points exchanging stability. 
Our present work thus serves also as a proof of principle, showing that such theories may be reliably probed via the bootstrap.

Let us finish by the outlook towards the bootstrap of conformal gauge theories. Despite recent progress in bootstrapping gauge theories, e.g. summarized in \cite{Poland:2022qrs}, obtaining an isolated island in parameter space corresponding to such theories still remains a very difficult task. 
The study of such theories may be improved by applying the methodology used in this paper, namely to attempt to track these theories from infinitesimal\footnote{In practice, in the bootstrap this would simply mean taking $\eps$ to be very small, such as e.g. $\varepsilon = 0.01$.} to finite and physically interesting values of a control parameter such as $1/N$ or $\varepsilon$.
This procedure may provide us with a more clear (and non-perturbative) picture for the spectrum of these complicated theories, which hopefully will give enough information about the various gaps needed to reliably isolate these theories in parameter space. 
See \cite{He:2021xvg} for some progress towards this direction.\footnote{See also \cite{Reehorst:2020phk,Li:2020bnb} for discussions of the manifestation of gauge-invariance in CFT spectra.}

\section*{Acknowledgements}
JH has received funding from the European Research Council (ERC) under the European
Union’s Horizon 2020 research and innovation programme (grant agreement no. 758903).
MR is supported by the Simons Foundation grant \#488659 (Simons collaboration on the non-perturbative bootstrap). 
The research work of SRK was initially supported by the Hellenic Foundation for
Research and Innovation (HFRI) under the HFRI PhD Fellowship grant
(Fellowship Number: 1026). The research work of SRK subsequently received funding from the European Research Council (ERC) under the European Union’s Horizon 2020 research and innovation programme (grant agreement no. 758903).
We thank the following people for useful conversations: Alessandro Vichi, Balt van Rees, Benoit Sirois, Bernardo Zan, Connor Behan, Gabriel Cuomo, Hugh Osborn, Ning Su, Slava Rychkov and Yuan Xin. Additionally, we thank an anonymous referee whose comments helped improve this manuscript.

The numerical computations were performed on the Metropolis cluster of the Crete Center for Quantum Complexity and Nanotechnology and on the Caltech High Performance Cluster, and were partially supported by a grant from the Gordon and Betty Moore Foundation.

\vspace{-12pt}
\begin{figure}[H]
  \flushright
  \includegraphics[scale=0.45]{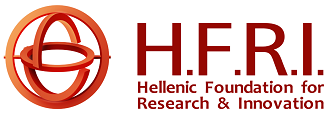}
\end{figure}

\vspace{-1.5cm}

\appendix

\section{Numerical setup}

For the numerics done in this work we set up various semi-definite programs (SDP) using the programs qboot \cite{Go:2020ahx} and Simpleboot \cite{simpleboot}. 
We integrated these with a python code to run the modified BFGS algorithm of \cite{Reehorst:2021ykw} to calculate the next point in the navigator optimization tasks. To solve the SDPs the arbitrary precision solver SDPB \cite{Simmons-Duffin:2015qma,Landry:2019qug} is used. 
To extract the spectrum we used the script \texttt{spectrum.py} \cite{spectrum} written originally for \cite{Komargodski:2016auf}, see also \cite{Simmons-Duffin:2016wlq} for more details. The script itself is an implementation of the extremal functional idea of \cite{El-Showk:2012vjm,El-Showk:2016mxr}. Note that this script discards operator for which it does not extract a positive OPE coefficient squared. For this study it was useful to also output these operators. Thus we used a modified version of \texttt{spectrum.py}.

The parameters we used when running SDPB can be found in table~\ref{tab:sdpb_parameters}.

\begin{table}[h]
	\begin{center}
{\small
\centering
			\begin{tabular}{ |c|c|c|c| } 
				\hline
				\texttt{$\Lambda$}  & 20 & 30\\ 
				\texttt{$\kappa$}
				 & 30 & 30\\ 
				\texttt{order}  & 60  & 60\\ 
				\texttt{spins} & \makecell{$\{0,1,2,\ldots,49\}\cup\{55,56,$\\$ 59, 60,64,65,69,70,74,$\\$75,79,80,84,85,89,90\}$} & \makecell{$\{0,1,2,\ldots,49\}\cup\{55,56,$\\$ 59, 60,64,65,69,70,74,$\\$75,79,80,84,85,89,90\}$} 
\\ 
				\texttt{precision} & 1100 & 1100   \\ 
				\texttt{dualityGapThreshold} & $10^{-30}$ & $10^{-30}$ \\ 
				\texttt{primalErrorThreshold} & $10^{-30}$ & $10^{-30}/10^{-60}$\\ 
				\texttt{dualErrorThreshold} & $10^{-30}$ & $10^{-30}/10^{-60}$\\ 
				\texttt{initialMatrixScalePrimal} & $10^{20}$ & $10^{20}/10^{40}$ \\ 
				\texttt{initialMatrixScaleDual} &  $10^{20}$  & $10^{20}/10^{40}$\\ 
				\texttt{feasibleCenteringParameter}  & 0.1 & 0.1 \\ 
				\texttt{infeasibleCenteringParameter} & 0.3 & 0.3 \\ 
				\texttt{stepLengthReduction} & 0.7 & 0.7\\ 
				\texttt{maxComplementarity} & $10^{100}$  & $10^{100}$ \\ 
				\hline
			\end{tabular}
		}
		\caption[]{ Parameters used to set up the SDPs, along with the \texttt{SDPB} parameters. The definition of these parameters can be found in \cite{Simmons-Duffin:2015qma}. In particular cases where we report two values, e.g $10^{-30}/10^{-60}$, this indicates that the bulk of our results were ran with the first value, however tests were performed to confirm that different values of this parameter did not change the results. In qboot we also use $n_{max}=500$. The parameters in the $\Lambda = 20$ column were used to explicitly check that our navigator minima lie in islands.
		}\label{tab:sdpb_parameters} 
	\end{center}
\end{table}

\section{Construction of the navigator function.}
\label{app:nav}
We start from the crossing equation for a $\mathbb{Z}_2$ symmetric CFT written in terms of sum rules:
\begin{equation}
\label{eq:crossing_original}
\vec{V}_{0,0} 
+ p_{\epsilon} \vec{V}_\epsilon (\theta)+ P_{T} \vec{V}_{T}
+ \sum_{(\Delta,\ell)\in S_+}  \Tr[P_{\Delta, \ell} \vec{V}_{+,\Delta,\ell}] 
+ \sum_{(\Delta,\ell)\in S_-} p_{\Delta,\ell} \vec{V}_{-,\Delta,\ell} = 0\,,
\end{equation}
where
\begin{equation} \quad \vec{V}_\epsilon(\theta) =\text{Tr}
\left[
\begin{pmatrix}c_\theta^2 &
c_\theta s_\theta\\ c_\theta s_\theta &  s^2_\theta\end{pmatrix}
\vec{V}_{+,\Delta_\epsilon,0}
+\begin{pmatrix}c_\theta^2 &0\\ 0 & 0\end{pmatrix}
\vec{V}_{-,\Delta_\sigma,0}
\right]. \label{Vepstheta}
\end{equation}
Here we follow the notation of \cite{Reehorst:2021ykw}.\footnote{These equations were first studied in \cite{El-Showk:2012cjh}. The fact that scanning over the OPE angle $\theta$ significantly improve the bootstrap bounds was first explored in \cite{Kos:2016ysd}.} For the definitions of $P_{\Delta,\ell}$, $p_{\Delta,\ell}$,  $\vec{V}_{+,\Delta_\epsilon,\ell}$,$\vec{V}_{-,\Delta_\sigma,\ell}$,  $\vec{V}_{0,0}$, $p_\epsilon$ and $\vec{V}_\epsilon$  see equations (2.14) to (2.17), (2.26) and (2.27) of that work. Here $P_T = \frac{\lambda^2_{\sigma\sigma T}}{\Delta_\sigma^2} \sim C_T^{-1}$ and $\vec{V}_T = \begin{pmatrix} \Delta_\sigma \\ \Delta_\epsilon \end{pmatrix} \cdot \vec{V}_{+,3,2} \cdot \begin{pmatrix} \Delta_\sigma \\ \Delta_\epsilon \end{pmatrix}$ see also \eqref{assumption4}. Finally, $c_\theta$ denotes $\cos{\theta}$ and $s_\theta$ denotes $\sin{\theta}$.  Our assumptions restrict the allowed operators that can appear in the sum above to
\begin{align}
S_+ =& \{(\Delta,0): \Delta\ge d\}\cup \{(\Delta,2): \Delta\ge d+1 \} \cup \{(\Delta,\ell): \ell =4,6,\ldots \text{ and }\Delta\ge d+ \ell-2\}\,, \label{Sp}
\\
S_- =& \{(\Delta,0): \Delta\ge d\}\cup \{(\Delta,\ell): \ell =1,2,3,\ldots \text{ and }\Delta\ge d+\ell-2\} \,.
\end{align}

Depending on the choice of $p=(\Delta_\sigma,\Delta_\epsilon,\theta)$ equation \eqref{eq:crossing_original} either does or does not have a solution. In order to efficiently find and classify allowed regions and their boundary we construct a navigator function $\mathcal{N}(\Delta_\sigma,\Delta_\epsilon,\theta)$. The key to finding a navigator function that has all the properties we require is a so called ``navigator-improved'' sum rule. By construction this sum rule has at least one solution with $\lambda=\lambda_*>0$ for all points and looks like this:
\begin{equation}
\label{eq:crossingIsingWithM}
\vec{V}_{0,0} + \lambda \vec{M} +
+ p_{\epsilon} \vec{V}_\epsilon (\theta)
+ \sum_{(\Delta,\ell)\in S_+}  \Tr[P_{\Delta, \ell} \vec{V}_{+,\Delta,\ell}] 
+ \sum_{(\Delta,\ell)\in S_-} p_{\Delta,\ell} \vec{V}_{-,\Delta,\ell} = 0\,.
\end{equation}
The vector $\vec{M}$ has to be chosen in order to guarantee the existence of a solution with $\lambda=\lambda_*>0$. Our choice of $M$ will be motivated by the observation that the generalized free field (GFF) would be a solution if not for the fact that we explicitly forbid contributions from operators with dimensions outside $S_+$ or $S_-$. Specifically, our assumptions forbid contributions from the GFF operators $\phi_1^2$, $\phi_2^2$, $\phi_1\phi_2$, $(T_{1 \mu\nu})'$ and $(T_{2 \mu\nu})'$ (where $\phi_1 = \sigma$ and $\phi_2 = \epsilon$). We can add these back to the equation through the $\vec{M}$ term to guarantee the existence of a solution with $\lambda=1$. This means taking
\begin{equation}
\vec{M}= \sum_{(\Delta,\ell)\in S^{\text{GFF}^*}_{+}}  \Tr[P^{\text{GFF}}_{\Delta, \ell} \vec{V}_{+,\Delta,\ell}] 
	+ \sum_{(\Delta,\ell)\in S^{\text{GFF}^*}_-} p^{\text{GFF}}_{\Delta,\ell} \vec{V}_{-,\Delta,\ell} = 0.
\end{equation} 
Here $S^{\text{GFF}^*}_{+}=\{(2\Delta_\sigma,0),(2\Delta_\epsilon,0),(2\Delta_\sigma+2,2)\}$ and 
$S^{\text{GFF}^*}_{-}=\{(\Delta_\sigma+\Delta_\epsilon,0)\}$. The OPE coefficients $P^{\text{GFF}}_{\Delta, \ell}$ and $p^{\text{GFF}}_{\Delta,\ell}$ have previously been computed in \cite{Fitzpatrick:2011dm,Fitzpatrick:2012yx,Karateev:2018oml}.

It is then easy to show that the navigator function
\begin{equation}
\N(\bf{p})=\min \lambda \quad \text{ such that equation \eqref{eq:crossingIsingWithM} has a solution}
\end{equation} 
 is bounded and has all the other properties we want as described in the main text.

\section{Auxiliary data}
\label{app:rawdata}

\subsection{Perturbative data}

In tables~\ref{tab:EvenEps} and \ref{tab:OddEps}, we display a collection of perturbative data for operators with dimension $\Delta\leqslant10$. Most of this data is available in computer-readable format in \cite{Henriksson:2022rnm}. The data for $\Z_2$-even scalars is displayed in tables~\ref{tab:EvenScalars} and \ref{tab:EvenScalarsOPE} in the main text.

\begin{table}[h]
\centering
\begin{tabular}{|c|c|c|lr|lr|l|}
\hline
$\ell$ & $i$ & $\O$ & \multicolumn{2}{c|}{$\Delta_\O$} & \multicolumn{2}{c|}{$\lambda^2_{\phi\phi\O}$} & \multicolumn{1}{c|}{$\lambda^2_{\phi^2\phi^2\O}$ }
\\\hline
\multirow{8}{*}{2}& 1 & $T_{\mu\nu}$ & $4-\varepsilon$ & & 
$\frac{d^2\Delta_\phi^2}{4(d-1)^2C_T}$ & &$\frac{d^2\Delta_{\phi^2}^2}{4(d-1)^2C_T}$ \rule{0pt}{4.5ex}    
\\
& 2 & $\de^2\phi^4$ & $6-\frac{5 \varepsilon}{9}$&  \cite{Kehrein:1994ff}  & $\frac{\varepsilon ^2}{1440}-\frac{\varepsilon^3}{1440}$ & \cite{Carmi:2020ekr,Bertucci:2022ptt} &  $\frac{8}{5}-\frac{52 \varepsilon }{27}$
 \rule{0pt}{3ex}    
\\
& 3 & $\square\de^2\phi^4$ &  $8-\frac{10 \varepsilon }{9}$ &  \cite{Kehrein:1994ff} & $\frac{\varepsilon^4}{2592000}$ & \cite{Bertucci:2022ptt} &  $\frac{1}{5}-\frac{1493 \varepsilon }{5400}$ 
\\
& 4 & $\de^2\phi^6$ &  $8+\frac{11 \varepsilon}{9}$ &  \cite{Kehrein:1994ff} &   && $O(\eps^2)$
\\
& 5 & $\square^2\de^2\phi^4$ &  $10-\frac{17\varepsilon}9$ & \cite{Henriksson:2022rnm} &   && $\frac1{999}$
\\
& 6 & $\square^2\de^2\phi^4$ &  $10-\frac{16\varepsilon}{15}$ & \cite{Henriksson:2022rnm} &   &&$\frac4{111}$
\\
& 7 & $\square\de^2\phi^6$ &  $10+\frac{4\varepsilon}9$ & \cite{Kehrein:1994ff} &   && $O(\eps^2)$
\\
& 8 & $\de^2\phi^8$ &  $10+\frac{13\varepsilon}3$ & \cite{Kehrein:1994ff} &   &&
\\\hline
\multirow{7}{*}{4}& 1 & $C_{\mu\nu\rho\sigma}$ & $6-\varepsilon +\frac{7 \varepsilon ^2}{540}$ & ($\varepsilon^4$)\cite{Derkachov:1997pf}& $\frac{1}{35}-\frac{1217 \varepsilon }{29400}
   $ & ($\varepsilon^4$)\cite{Alday:2017zzv} &  $\frac{4}{35}-\frac{1831 \varepsilon }{22050}$
\\
& 2 & $(\de^4\phi^4)_1$ &  $8-\frac{14 \varepsilon }{9}$ & \cite{Kehrein:1994ff} & $\frac{\varepsilon ^2}{419580}$ & \cite{Alday:2017zzv,Henriksson:2022rnm} & $\frac{125}{2331}$
\\
& 3 & $(\de^4\phi^4)_2$ & $8-\frac{11 \varepsilon }{15}$ & \cite{Kehrein:1994ff} & $\frac{\varepsilon^2}{23976}$   & \cite{Alday:2017zzv,Henriksson:2022rnm} &  $\frac{8}{37}$
\\
& 4 & $(\square\de^4\phi^4)_1$ &  $10-\frac{\sqrt{697} +131  }{90}\varepsilon$& \cite{Henriksson:2022rnm} &&& $0.00201$
\\
& 5 & $(\square\de^4\phi^4)_2$& $10-\frac{-\sqrt{697} +131  }{90}\varepsilon$ & \cite{Henriksson:2022rnm} &&& $0.03694$  
\\
& 6 & $(\de^4\phi^6)_1$& $10+\frac{13 \varepsilon }{15}$& \cite{Kehrein:1994ff}&&& $O(\eps^2)$
\\
& 7 & $(\de^4\phi^6)_2$& $10+0\eps$ & \cite{Kehrein:1994ff} &&& $O(\eps^2)$
\\\hline
\multirow{4}{*}{6}& 1 & $\mathcal J_6$ & $8-\varepsilon +\frac{\varepsilon ^2}{63}$ & ($\varepsilon^4$)\cite{Derkachov:1997pf} & $\frac{1}{462}-\frac{49807 \varepsilon }{12806640}$ & ($\varepsilon^4$)\cite{Alday:2017zzv} &  $\frac{2}{231}-\frac{23011 \varepsilon
   }{3201660}$
\\
& 2 & $(\de^6\phi^4)_1$   &  $10-1.50342 \varepsilon$ &\cite{Kehrein:1994ff}  &   $1.67\cdot 10^{-7}\varepsilon^2$ &\cite{Alday:2017zzv,Henriksson:2022rnm} & $O(1)$
\\ 
& 3 & $(\de^6\phi^4)_2$ &  $10-1.32928 \varepsilon$ & \cite{Kehrein:1994ff} & $0.76\cdot 10^{-7}\varepsilon^2$ &\cite{Alday:2017zzv,Henriksson:2022rnm} &$O(1)$
\\
& 4 & $(\de^6\phi^4)_3$ & $10-0.82126\varepsilon$ & \cite{Kehrein:1994ff} & $27.5 \cdot 10^{-7}\varepsilon^2$ &\cite{Alday:2017zzv,Henriksson:2022rnm} & $O(1)$
\\\hline
8& 1 & $\mathcal J_8$ & $10-\varepsilon+\frac{11 \varepsilon ^2}{648} $ & ($\varepsilon^4$)\cite{Derkachov:1997pf} & $\frac1{6435} +\ldots$ 
  &($\varepsilon^4$)\cite{Alday:2017zzv}  & $\frac4{6435} +\ldots$ 
\\\hline
\end{tabular}
\caption{Perturbative results for $\Z_2$-even operators with $\Delta\leqslant10$ of even spin $\ell>0$.
The OPE coefficients $\lambda^2_{\phi^2\phi^2\O}$ can all be extracted from the correlator given in \cite{Henriksson:2020jwk}, see also \cite{Henriksson:2022rnm,Bertucci:2022ptt} For $\lambda^2_{\phi^2\phi^2\mathcal J_\ell}$, with $\ell=4,6,\ldots$, order $\epsilon^2$ results are also available in \cite{Bertucci:2022ptt}. For entries that are empty, it is not known to what order in $\eps$ the OPE coefficients appear.
}\label{tab:EvenEps}
\end{table}

\begin{table}
\centering
\begin{tabular}{|c|c|c|lr|lr|}
\hline
$\ell$ & $i$ & $\O$ & \multicolumn{2}{c|}{$\Delta_\O$} & \multicolumn{2}{c|}{$\lambda^2_{\phi\phi^2\O}$} 
\\\hline
\multirow{5}{*}{0}& 1 & $\phi$ &   $1-\frac{\epsilon }{2}+\frac{\epsilon
   ^2}{108}$   & ($\epsilon^8$)\cite{SchnetzUnp}  &  $2-\frac{2 \epsilon }{3}-\frac{34 \epsilon^2}{81}$   & ($\epsilon^4$)\cite{Carmi:2020ekr}
\\
& 2 & $\phi^5$ & $5+\frac{5 \epsilon }{6}-\frac{685 \epsilon ^2}{324}$  & ($\epsilon^3$)\cite{Zhang1982}   & $\frac{5 \epsilon ^2}{108}$ & \cite{Codello:2017qek}
\\
& 3 & $\phi^7$ & $7+\frac{7 \epsilon }{2}-\frac{959 \epsilon
   ^2}{108}$ & \cite{Derkachov:1997gc} &&
\\
& 4 & $\square^2\phi^5$ & $9-\frac{5 \epsilon
   }{18}$ &  \cite{Kehrein:1994ff}  &&
\\
& 5 & $\phi^9$ & $9+\frac{15 \epsilon }{2}-\frac{821 \epsilon^2}{36}$  & \cite{Derkachov:1997gc} &&
\\\hline
1 & 1 & $\square^2\de\phi^5$  &  $10-\frac{7 \epsilon }{9}$ &  \cite{Kehrein:1994ff}  &&
\\\hline
\multirow{4}{*}{2}& 1 & $\de^2\phi^3$  & $5-\frac{17 \eps}{18}+\frac{161 \epsilon^2}{972}$  & \cite{Bertucci:2022ptt} & $\frac{1}{2}-\frac{199 \eps}{432}$ & \cite{Bertucci:2022ptt}
\\
& 2 & $\de^2\phi^5$ & $7+\frac{\eps}{6}$ &  \cite{Kehrein:1994ff}  &&
\\
& 3 & $\square\de^2\phi^5$ & $9-\frac{\eps}{2}$ & \cite{Kehrein:1994ff} &&
\\
& 4 & $\de^2\phi^7$ & $9+\frac{47 \eps}{18}$  & \cite{Kehrein:1994ff} &&
\\\hline
\multirow{4}{*}{3}& 1 & $\de^3\phi^3$ & $6-\frac{4 \eps}{3}+\frac{17 \eps^2}{2592}$& \cite{Bertucci:2022ptt}& $\frac{2}{35}-\frac{911 \eps}{14700}$ & \cite{Bertucci:2022ptt}
\\
& 2 & $\de^3\phi^5$ &  $8-\frac{\eps}{3}$ &  \cite{Kehrein:1994ff}  &  &
\\
& 3 & $\square\de^3\phi^5$ & $10-\frac{10 \eps}{9}$ &  \cite{Kehrein:1994ff}  &  &
\\ 
& 4 & $\de^3\phi^7$ & $10+2\eps$ &  \cite{Kehrein:1994ff}  &  &
\\\hline
\multirow{3}{*}{4}& 1 & $\de^4\phi^3$ &$7-\frac{31 \epsilon }{30}+\frac{3559 \eps^2}{20250}$ & \cite{Bertucci:2022ptt} &$\frac{1}{18}-\frac{13943 \eps }{194400}$&\cite{Bertucci:2022ptt}
\\
& 2 & $(\de^4\phi^5)_1$ &$9-\frac{17 \eps   }{18}$ &  \cite{Kehrein:1994ff}  &&
\\
& 3 & $(\de^4\phi^5)_2$ & $9-\frac{\eps}{10}$ &  \cite{Kehrein:1994ff}  &&
\\\hline
\multirow{3}{*}{5}& 1 & $\de^5\phi^3$ & $8-\frac{23 \eps}{18}+\frac{61 \eps^2}{1620}$ &\cite{Bertucci:2022ptt}& $\frac{2}{231}-\frac{378163 \eps}{28814940}$ &\cite{Bertucci:2022ptt}
   \\
& 2 &  $(\de^5\phi^5)_1$  &  $10-\frac{5 \eps}{3}$ &  \cite{Kehrein:1994ff}  &&
   \\
& 3 &  $(\de^5\phi^5)_2$    &  $10-\frac{7 \eps}{18}$ &  \cite{Kehrein:1994ff}  &&
\\\hline
\multirow{2}{*}{6}& 1 & $(\de^6\phi^3)_1$ & $9-\frac{3 \eps}{2}$ & \cite{Kehrein:1992fn} &&
\\
& 2 & $(\de^6\phi^3)_2$ & $9-\frac{15 \epsilon
   }{14}+\frac{48799 \eps^2}{277830}$&\cite{Bertucci:2022ptt}&$\frac{3}{572}-\frac{8019289 \eps}{961920960}$&\cite{Bertucci:2022ptt}
 \\\hline
 7 & 1 & $\de^7\phi^3$  & $10-\frac{5 \eps}{4}+\frac{4297 \eps^2}{80640}$ &\cite{Bertucci:2022ptt}  & $\frac{2}{2145}-\frac{2625239 \eps }{1545944400}$&\cite{Bertucci:2022ptt}  
\\\hline
\end{tabular}
\caption{Perturbative results for $\Z_2$-odd operators with $\Delta \leqslant10$.}\label{tab:OddEps}
\end{table}

\subsection{Navigator minima}
\label{sec:navigatorminima}

In table~\ref{tab:navigatorminima} we show the values for the navigator minima in the different spacetime dimensions studied.

\begin{table}[h]
\centering
\begin{tabular}{|c|c|c|c|}
\hline
$d$ & $\Delta_{\sigma}|_{\mathrm{min}}$ & $\Delta_{\epsilon}|_{\mathrm{min}}$ & $\theta|_{\mathrm{min}}$
\\\hline
2.6 & 0.3440987308 & 1.219622499 & 0.8616022236 \\ 
2.7 & 0.3860503791 & 1.265155812 & 0.8968882905 \\ 
2.8 & 0.4291188478 & 1.312594686 & 0.9253317624 \\ 
2.9 & 0.4731867209 & 1.361792983 & 0.9490037256 \\ 
3. & 0.518148953 & 1.412627358 & 0.9692651416 \\ 
3.1 & 0.5639124084 & 1.465006065 & 0.9870522563 \\ 
3.2 & 0.6103917163 & 1.518841816 & 1.003021129 \\ 
3.3 & 0.6575105206 & 1.574075519 & 1.017657414 \\ 
3.4 & 0.7051992474 & 1.630668115 & 1.031330689 \\ 
3.5 & 0.7533940071 & 1.688601476 & 1.044333198 \\ 
3.6 & 0.8020356597 & 1.747882413 & 1.056907458 \\ 
3.7 & 0.8510688458 & 1.808547471 & 1.069266184 \\ 
3.8 & 0.9004410692 & 1.870674384 & 1.081610779 \\ 
3.9 & 0.9501016191 & 1.934405213 & 1.094152721 \\ 
3.95 & 0.9750242935 & 1.966945908 & 1.100574515
\\\hline
\end{tabular}
\caption{Table of navigator minima.
}
\label{tab:navigatorminima}
\end{table}

\subsection{Operators for fit of level repulsion plot}
\label{app:opsforfit}

For the fit in section~\ref{sec:levelRepRes}, we make use the points
\begin{align}
d&=2.6: \qquad \Delta\in\{6.96815028,\, 7.30348360\},
\nonumber
\\
d&=2.7: \qquad \Delta \in\{7.04476307,\, 7.20549735\},
\nonumber
\\
d&=2.8: \qquad \Delta \in\{7.04351005,\, 7.19424952\},
\\
d&=2.9: \qquad \Delta \in\{6.98086197,\, 7.24126426\},
\nonumber
\\
d&=3.0: \qquad \Delta \in\{6.90348236,\, 7.29743289\}.
\nonumber
\end{align}

\section{Extremal functional method}
\label{sec:EFM}

\begin{figure}
	\centering
	\includegraphics[width=0.85\textwidth]{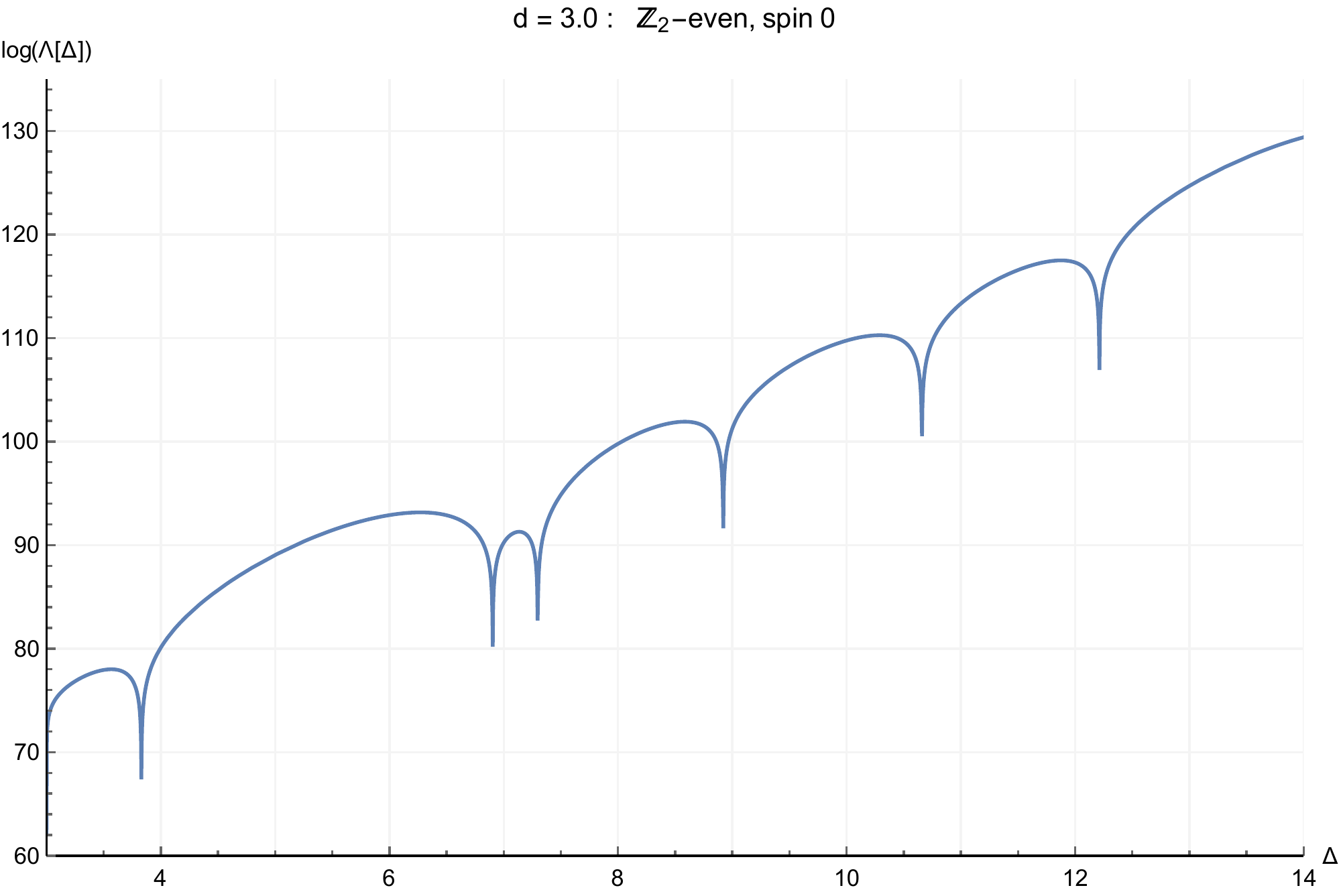}
	\caption{Extremal functional $\Lambda(\Delta)$ for $\mathbb Z_2$-even scalars of dimensions $\Delta$ at $d=3.0$ dimensions.}\label{fig:EFMeven0}
\end{figure}

A key step in our analysis is the use of the extremal functional method (EFM), developed in \cite{Poland:2010wg,El-Showk:2012vjm,El-Showk:2016mxr}. For demonstration purposes we show what the action of such a functional looks like on a sum rule. More specifically, in figure~\ref{fig:EFMeven0}, we show the determinant of the extremal functional acting on the $\mathbb{Z}_2$-even spin-$0$ sector in $d=3$, using the parameters from the second column of table~\ref{tab:sdpb_parameters}, at the point in parameter space referred to in the caption of table~\ref{table:improvement}.

From figure~\ref{fig:EFMeven0} we read off the following minima,
\begin{align}
\label{eq:E-0-30}
\Delta\in & \{3.82977589284,\,6.90348236615,\,7.29743289116,\ 
\nonumber
\\
& 8.92058769366,\,10.6590253446,\,12.2118684512,\,14.9182183792,\ldots \}.
\end{align}
Alternatively, if we use the automatic reading of \texttt{spectrum.py}, we find
\begin{equation}
\label{eq:E-0-30-spectrumPy}
\Delta\in
\{3.82977589284,\,6.90348236615,\,10.6590253446,\,12.2118684512,\ldots\}.
\end{equation}
It is thus clear that this misses operators,  in this case at $\Delta=7.2974$ and $\Delta=8.9206$. In fact \texttt{spectrum.py} is able to detect these minima, but it discards them because solving the primal solution for the OPE coefficients it finds a negative squared OPE coefficient. 

We believe the inaccuracy in the OPE coefficients to be due to a numerical instability. Hence, in order to be able to detect all operators, we read of the minima of the determinants of the functionals in Mathematica. Or, equivalently, using a  modified version of \texttt{spectrum.py} that does not discard these operators.

\subsection{Systematic errors}
\label{sec:systematicErr}

For the case of $d=3$, we can compare with the values in \cite{Simmons-Duffin:2016wlq}. For $\Z_2$-even scalars above $\epsilon$, that work found three stable operators, and some other visible features. In summary
\begin{equation}
\Delta\in\{3.82968(23) ,\,6.8956(43),\,7.2535(51),\,  \sim 8.55,\,\sim10.45,\,\sim11.6\}.
\end{equation}
We can see that for operators with $\tau\gtrsim 5$, our values are consistently above those of \cite{Simmons-Duffin:2016wlq}. It is likely that the true operator dimensions are below those of both works. In table~\ref{table:improvement} we evaluate the minima of the extremal functional for various values of $\Lambda$. One may also compare with the values of \cite{Reehorst:2021hmp}, which obtained $\Delta_{\epsilon'}=3.82951(61)$ with a rigorous error bar; our result is $\Delta_{\epsilon'}=3.82978$. See also, \eqref{opecomparisons} and \eqref{opecomparisons2} for a comparison of the leading OPE coefficients against values with rigorous error bars.

\begin{table}
	\begin{tabular}{|l|llllll|}
		\hline
		&   $\Delta_{\epsilon'}$  &   $\Delta_{\epsilon''}$  &   $\Delta_{\epsilon'''}$  &   $\Delta_{\epsilon''''}$  &   $\Delta_{\epsilon'''''}$  &   $\Delta_{\epsilon''''''}$ 
		\\\hline
		$\Lambda=15$ & $3.82994 $&$ 7.20245  $&$ 7.64939  $&$ 9.23416  $&$ 13.0288 $&
		\\
		$\Lambda=21$ &  $3.83012  $&$ 6.97177  $&$ 7.25821  $&$ 8.64518  $&$ 11.3409  $&$12.9575$
		\\
		$\Lambda=30$ & $ 3.82978 $&$ 6.90348  $&$ 7.29743  $&$ 8.92059  $&$ 10.6590  $&$ 12.2119$
		\\
		$\Lambda=43$ \cite{Simmons-Duffin:2016wlq} &  $3.82968(23)$&$6.8956(43)$&$7.2535(51)$&$  \sim 8.55$&$\sim10.45$&$\sim11.6$.
		\\\hline
	\end{tabular}
	\caption{Zeros of the functional for $\mathbb Z_2$-even scalars in $d=3$ at different $\Lambda$. The first three lines were computed for the specific point ($\Delta_\phi$ , $\Delta_{\phi^2}$ , $\theta$) $=$ ($0.518148953$ , $1.412627358$ , $0.9692651416$). The last line corresponds to the averaged value of \cite{Simmons-Duffin:2016wlq}.}\label{table:improvement}	
\end{table}

{
\bibliographystyle{JHEP}
\bibliography{biblio}
}

\end{document}